\documentstyle[aps,floats,epsfig]{revtex}

\begin{document}

\draft

\title{Dynamic Cosmic Strings II: Numerical evolution of excited Cosmic
       Strings}

\author{
U Sperhake\thanks{E-mail: U.Sperhake@maths.soton.ac.uk},
K R P Sj\"odin\thanks{E-mail: K.R.Sjodin@maths.soton.ac.uk}  
and J A Vickers\thanks{E-mail: J.A.Vickers@maths.soton.ac.uk}}

\address{
Faculty of Mathematical Studies, \\
University of Southampton, \\ 
Southampton, S017 1BJ, U.K.
}

\date{31 March 2000}

\maketitle


\begin{abstract}
An implicit, fully characteristic, numerical scheme for solving
the field equations of a cosmic string coupled to gravity is
described. The inclusion of null infinity as part of the numerical
grid allows us to apply suitable boundary conditions on the metric
and matter fields to suppress unphysical divergent solutions.
The code is tested by comparing the results with exact
solutions, checking that static cosmic string initial data
remain constant when evolved and undertaking a time dependent
convergence analysis of the code. It is shown that the code is
accurate, stable and exhibits clear second order convergence. The code
is used to analyse the interaction between a Weber--Wheeler pulse of
gravitational radiation with the string. The interaction causes the
string to oscillate at frequencies proportional to the
masses of the scalar and vector fields of the string. After the pulse
has largely radiated away the string
continues to ring but the oscillations slowly decay and eventually the
variables return to their static values.
\end{abstract}


\section{Introduction}
This is the second paper in a series devoted to the study of dynamic
cosmic strings. In the previous paper \cite{SSV}, henceforth referred
to as paper I, we derived the equations of motion for a time dependent
cylindrically symmetric cosmic string coupled to a gravitational field
with two degrees of freedom. The treatment involved using a Geroch
decomposition to reformulate the problem in terms of matter fields on 
a $2+1$ dimensional spacetime and two geometrical variables $\nu$ and
$\tau$ which describe the gravitational degrees of freedom. Unlike the
original cylindrical metric the reduced $2+1$ spacetime is
asymptotically flat and this allows us to implement a numerical
treatment of the field equations which includes null infinity as part
of the grid. In paper I we presented a Cauchy-characteristic matching
(CCM) code that reproduced the results of several vacuum solutions
with both one and two degrees of freedom but did not perform entirely
satisfactorily in the presence of matter. We have therefore developed
a second implicit, purely characteristic code.
%
%
The numerical details, the testing and the convergence analysis of this
implicit code will be described in this paper.
We also give a detailed analysis of
the interaction between the string and a pulse of gravitational
radiation.

After briefly describing the variables used for the coupled system in
section II the details of the numerical scheme are presented in
section III. We begin by describing the relaxation method we use for the 
much simpler problem of a static cosmic string in Minkowski spacetime and 
demonstrate how this approach leads naturally to the implicit scheme
used to solve the general case of a dynamical cosmic string coupled to
gravity. The testing of the code is described in section IV. This
involves comparing it with exact solutions, checking that static
cosmic string initial data remain static when they are evolved, and
undertaking a time dependent convergence analysis of the code. We are
able to show that the code is accurate, stable and shows clear second
order convergence. In section V we analyse the interaction between an
initially static cosmic string and a Weber--Wheeler type pulse of
gravitational radiation. We show that this interaction causes the
string fields $S$ and $P$ to oscillate and examine how this
oscillation depends upon both the width and amplitude of the pulse and
also upon the constants $e$, $\lambda$ and $\eta$. The key
result is that the frequencies are essentially independent of the
nature of the Weber--Wheeler wave but are proportional to the
masses $m_S$ and $m_P$ associated with the scalar field $S$ and the
vector field $P$. The string continues to ring after the gravitational
pulse has largely radiated away, but the oscillations slowly decay and
the variables eventually return to their static values. It is only
possible to observe this effect because of the long term stability of
the code. Finally in section VI we discuss our results and outline
future work. 

\section{The variables and field equations}

We begin with a brief summary of the variables used to describe the
spacetime and the string. In cylindrical polar coordinates
$(t,\rho,\phi,z)$ we may write the line element for a cylindrically
symmetric spacetime with two degrees of freedom as a modified version
of that given by Jordan, Ehlers, Kundt and Kompaneets \cite{JEK}, \cite{Komp}
\begin{equation} 
ds^2 = e^{2(\gamma-\psi)}(dt^2 - d\rho^2) - \rho^2e^{-2\psi}d\phi^2 
- e^{2(\psi+\mu)}(\omega d\phi+dz)^2,
\label{4d}
\end{equation}
where $\psi$, $\omega$, $\mu$ and $\gamma$ are functions of $t$, $\rho$. 
As shown in paper I, in order to work with an asymptotically flat
spacetime it is convenient to make a Geroch decomposition \cite{Geroch}
in which the 4-dimensional metric is replaced by a 3-dimensional one
and two auxiliary scalar fields. These scalar fields are the norm of the
axial Killing vector $\nu$ and the Geroch potential $\tau$ and are
related to $\psi$ and $\omega$ by equations (15) and (16) of paper
I, while the conformal 3-dimensional line element is given by
\begin{equation}
d{\tilde\sigma}^2=e^{2(\gamma+\mu)}(dt^2-d\rho^2)-\rho^2e^{2\mu}d\phi^2.
\end{equation}
The cosmic string is described by a complex scalar field $\Phi$
coupled to a U(1) gauge field $A_\mu$, which in cylindrical symmetry
can be written as \cite{SSV}
\begin{eqnarray}
\Phi&=&\frac{1}{\sqrt 2}S(t,\rho)e^{i\phi}, \\
A_\mu&=&\frac{1}{e}[P(t,\rho)-1]\nabla_\mu\phi .
\end{eqnarray}
It is also helpful to introduce rescaled coupling constants and
variables given by
\begin{eqnarray}
    X &=& \frac{S}{\eta},\\
    \alpha &=& \frac{e^2}{\lambda},\\
    r &=& \sqrt\lambda \eta\rho, \\
    {\tilde t} &=& \sqrt\lambda \eta t. 
\end{eqnarray}
Here $\eta$ is the vacuum expectation value of the scalar field while
$\alpha$ represents the relative strength of the coupling between the
scalar and vector field given by $e$, compared to the self-coupling of
the scalar field given by $\lambda$. Critical coupling, for which the
masses of the scalar and vector fields are equal, is given by
$\alpha=8$ \cite{shellard}.

The field equations for a cosmic string coupled to gravity were 
derived in paper I in terms of Cauchy coordinates [Paper I
(26)--(33)] and  also in terms of compactified characteristic
coordinates [Paper I (43)--(49)]. However the fully characteristic
numerical scheme also requires characteristic equations in the inner
region. In terms of the retarded time $u=\tilde t-r$ and the
radius $r$ the field equations are given by
\begin{eqnarray}
  \Box \nu &=& \nu_{,r} \mu_{,r} + \frac{\tau_{,r}^2-\nu_{,r}^2}{\nu}
        -\nu_{,u} \mu_{,r} - \nu_{,r} \mu_{,u}
        +2\frac{\nu_{,u} \nu_{,r} - \tau_{,u} \tau_{,r}}{\nu}
        +8\pi \eta^2\left[ 2e^{2(\gamma+\mu)} (X^2-1)^2
        +e^{-2\mu} \nu^2
        \frac{2P_{,u} P_{,r} - P_{,r}^2}{\alpha r^2} \right], \label{nuur} 
        \\[10pt]
  \Box \tau &=& \tau_{,r}\mu_{,r} - 2\frac{\tau_{,r} \nu_{,r}}{\nu}
        - \tau_{,u} \mu_{,r}-\tau_{,r} \mu_{,u}
        + 2\frac{\tau_{,r} \nu_{,u} +\tau_{,u} \nu_{,r}}{\nu}, \\[10pt]
  \Box \mu &=& \mu_{,r}^2 +\frac{\mu_{,r}}{r}
        - \frac{\mu_{,u}}{r} -2\mu_{,u} \mu_{,r} +8\pi\eta^2 \left[ 2
        \frac{e^{2(\gamma + \mu)}}{\nu} (X^2-1)^2 + e^{2\gamma}
        \frac{X^2P^2}{r^2} \right], \\[10pt]
  0 &=& 2\gamma_{,r} +2r\gamma_{,r} \mu_{,r} -r\mu_{,rr} + r\mu_{,r}^2
        -\frac{r}{2\nu^2} (\tau_{,r}^2 + \nu_{,r}^2) - 8\pi \eta^2 \left[
        r X_{,r}^2 + \frac{1}{\alpha} e^{-2\mu} \nu \frac{P_{,r}^2}{r} \right],
        \label{gammar} \\ [10pt]
  \Box P &=& -2\frac{P_{,r}}{r} +2\frac{P_{,u}}{r} - P_{,r} \mu_{,r}
        + P_{,r}\frac{\nu_{,r}}{\nu} + P_{,r} \mu_{,u} + P_{,u} \mu_{,r}
        - \frac{P_{,r} \nu_{,u} + P_{,u} \nu_{,r}}{\nu}
        - \alpha \frac{e^{2(\gamma + \mu)}}{\nu} PX^2,
        \label{pur} \\[10pt]
  \Box X &=& X_{,r} \mu_{,r} - X_{,u} \mu_{,r} - X_{,r} \mu_{,u}
        - 4  \frac{e^{2(\gamma + \mu)}}{\nu} X(X^2-1)
        - e^{2\gamma} \frac{XP^2}{r^2}, \label{xur}
\end{eqnarray}
where $\Box$ represents the flat-space d'Alembert operator
\begin{equation}
  \Box = 2\frac{\partial^2}{\partial u \partial r} 
         - \frac{\partial^2}{\partial r^2} - \frac{1}{r} \left(
         \frac{\partial}{\partial r} - \frac{\partial}{\partial u} \right).
\end{equation}
Note that there are two further Einstein equations which are not used in
the numerical scheme as they are a consequence of the above equations
and their derivatives. These equations are only used to provide a check on
the numerical accuracy of the code.

\section{Numerical methods}

In order to solve the above field equations we have developed two
independent codes. The first is based on
the Cauchy-characteristic matching code of Dubal et al.
\cite{DDC} and d'Inverno et al. \cite{DDS}. This code performs well in
the absence of matter and has been used in paper I to study several
cylindrically symmetric vacuum solutions. In paper I we have also given
details of the convergence analysis and described
the modifications with respect to \cite{DDC} that lead to long term stability
with both gravitational degrees of freedom present. However the CCM code
performed less satisfactorily in the evolution of the cosmic string. This
is due to the existence of
unphysical solutions to the evolution equations (\ref{nuur})--(\ref{xur}) which
diverge exponentially as $r\rightarrow \infty$. Controlling the time evolution
near null infinity by means of a bump function enabled us to select the
physical solutions with regular behavior at $I^+$, but the bump function
itself introduced noise which eventually gave rise to instabilities.
We therefore implemented a second implicit,
purely characteristic, code which
allowed us to directly apply boundary conditions at the origin as well
as null infinity and thus suppress diverging solutions. It is interesting
that this problem is already present in the calculation of the static
cosmic string in Minkowski spacetime. We will,
therefore, first describe the numerical scheme used in the static Minkowskian
case where the equations are fairly simple.  We then present the modifications
necessary for the static and dynamic case coupled to the gravitational
field.

\subsection{The static cosmic string in Minkowski spacetime}

In (\ref{nuur})--(\ref{xur})
we set the metric variables to their Minkowskian values and all time
derivatives to zero to obtain the equations for the static cosmic string in
Minkowski spacetime (cf.\ \cite{garfinkle})
\begin{eqnarray}
  r\frac{d}{dr} \left( r^{-1} \frac{dP}{dr} \right) &=& \alpha X^2 P ,
        \label{mink_Prr} \\[15pt]
  r\frac{d}{dr}\left( r\frac{dX}{dr} \right) &=& X \left[ P^2 + 4r^2 (X^2-1)
        \right] . \label{mink_Xrr}
\end{eqnarray}
The boundary conditions are \cite{garfinkle}
\begin{eqnarray}
  & P(0) = 1,  & \qquad \qquad \qquad \lim_{r \rightarrow \infty} P(r) = 0, 
    \nonumber \\[10pt]
  & X(0) = 0,  & \qquad \qquad \qquad \lim_{r \rightarrow \infty} X(r) = 1.
    \label{mink_bound}
\end{eqnarray}
\begin{figure}[t]
\centering

\begin{picture}(0,0)%
\epsfig{file=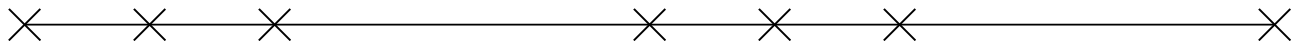}%
\end{picture}%
\setlength{\unitlength}{3947sp}%
\begingroup\makeatletter\ifx\SetFigFont\undefined%
\gdef\SetFigFont#1#2#3#4#5{%
  \reset@font\fontsize{#1}{#2pt}%
  \fontfamily{#3}\fontseries{#4}\fontshape{#5}%
  \selectfont}%
\fi\endgroup%
\begin{picture}(6312,2028)(976,-1606)
\put(3226,-436){\makebox(0,0)[lb]{\smash{\SetFigFont{11}{13.2}{\rmdefault}{\mddefault}{\updefault}...}}}
\put(2401,-436){\makebox(0,0)[lb]{\smash{\SetFigFont{11}{13.2}{\rmdefault}{\mddefault}{\updefault}3}}}
\put(1801,-436){\makebox(0,0)[lb]{\smash{\SetFigFont{11}{13.2}{\rmdefault}{\mddefault}{\updefault}2}}}
\put(976,-436){\makebox(0,0)[lb]{\smash{\SetFigFont{11}{13.2}{\rmdefault}{\mddefault}{\updefault}$k$=1}}}
\put(976,-61){\makebox(0,0)[lb]{\smash{\SetFigFont{11}{13.2}{\rmdefault}{\mddefault}{\updefault}$r$=0}}}
\put(3976,-1186){\makebox(0,0)[lb]{\smash{\SetFigFont{11}{13.2}{\rmdefault}{\mddefault}{\updefault}$n_1$+1}}}
\put(4051,-436){\makebox(0,0)[lb]{\smash{\SetFigFont{11}{13.2}{\rmdefault}{\mddefault}{\updefault}$n_1$}}}
\put(4051,-61){\makebox(0,0)[lb]{\smash{\SetFigFont{11}{13.2}{\rmdefault}{\mddefault}{\updefault}$r$=1}}}
\put(4051,-1561){\makebox(0,0)[lb]{\smash{\SetFigFont{11}{13.2}{\rmdefault}{\mddefault}{\updefault}$y$=1}}}
\put(4651,-1186){\makebox(0,0)[lb]{\smash{\SetFigFont{11}{13.2}{\rmdefault}{\mddefault}{\updefault}$n_1$+2}}}
\put(5251,-1186){\makebox(0,0)[lb]{\smash{\SetFigFont{11}{13.2}{\rmdefault}{\mddefault}{\updefault}$n_1$+3}}}
\put(5701,314){\makebox(0,0)[lb]{\smash{\SetFigFont{11}{13.2}{\rmdefault}{\mddefault}{\updefault}outer region}}}
\put(2101,314){\makebox(0,0)[lb]{\smash{\SetFigFont{11}{13.2}{\rmdefault}{\mddefault}{\updefault}inner region}}}
\put(6901,-1186){\makebox(0,0)[lb]{\smash{\SetFigFont{11}{13.2}{\rmdefault}{\mddefault}{\updefault}$n_1$+$n_2$}}}
\put(7051,-1561){\makebox(0,0)[lb]{\smash{\SetFigFont{11}{13.2}{\rmdefault}{\mddefault}{\updefault}$y$=0}}}
\put(6376,-1186){\makebox(0,0)[lb]{\smash{\SetFigFont{11}{13.2}{\rmdefault}{\mddefault}{\updefault}...}}}
\end{picture}

\vspace{0.5cm}
\caption{The combined grid of the inner and the outer region. Note that
         both grid points, $n_1$ and $n_1+1$, correspond to the position
         $r=1 \Leftrightarrow y=1$.
         These points form the interface of the code and facilitate
         transformation of the variables from the coordinate system $r$
         into that of $y$.}
\label{grid}
\end{figure}
In order to cover the whole spacetime with a finite coordinate range, 
we divide the
computational domain into two regions. In the inner region ($0\le r\le 1$) we
use the coordinate $r$, while in the outer region we introduce the compactified
radius
\begin{eqnarray}
  y &=& \frac{1}{\sqrt{r}},
\end{eqnarray}
which covers the range $1 \ge y \ge 0$ corresponding to the region 
$r \ge 1$. It is also useful to combine $r$ and $y$ into a
single radial variable $w$ defined by
\begin{eqnarray}
  w = \left\{\parbox{40mm} {$r \qquad \qquad \,\, {\rm for} 
                             \,\,\,\, 0\le r \le 1 \\
                             3-2/\sqrt{r} \hspace{0.28cm} {\rm for} \,\,\,\, 
                             r>1$,}\right. 
      \label{w}
\end{eqnarray}
so that the region $0 \leq w < 3$ corresponds to 
$0 \leq r < \infty$ and infinity is mapped to $w=3$ \cite{SSV}.

In terms of the variable $y$, (\ref{mink_Prr}) and
(\ref{mink_Xrr}) take the form
\begin{eqnarray}
  y \frac{d}{dy} \left( y^5\frac{dP}{dy} \right) &=& 4\alpha X^2P,\\
  y \frac{d}{dy} \left( y \frac{dX}{dy} \right)  &=& 
  4X\left[P^2+ 4\frac{(X^2-1)}{y^4}\right].
\end{eqnarray}
The number of grid points in each region may differ, but
each half-grid is uniform. Thus we use
a total of $N:=n_1 + n_2$ grid points where the points labelled $n_1$ and
$n_1+1$ both correspond to the position $r=1=y$. 
The points $n_1$, $n_1+1$ form the interface between the two regions
(see Figure \ref{grid}). One point will contain
the variables in terms of $r$, the other in terms of $y$. 
%
%
With the computational grid covering the whole spacetime, we now face a
two point boundary value problem. Due to the existence of unphysical solutions
diverging at $y=0$ shooting methods turned out to be unsuitable for solving
this problem. On the other hand numerical relaxation, 
as described in \cite{relax} for example, 
allows us to directly control the behavior of $P$ and $X$ at infinity. 
The form of equations (\ref{mink_Prr}),\ (\ref{mink_Xrr}) suggests that
in order to write them as a first order system we should introduce the 
auxiliary variables $Q=r^{-1}P_{,r}$ and $R=rX_{,r}$. The equations may then be
written in the form
\begin{eqnarray}
  P_{,r} &=& rQ, \\[10pt]
  X_{,r} &=& \frac{R}{r}, \\[10pt]
  Q_{,r} &=& \alpha \frac{PX^2}{r}, \\[10pt]
  R_{,r} &=& X\left[\frac{P^2}{r} + 4r(X^2-1)\right] .
\end{eqnarray}
The corresponding equations in the outer region are given by
\begin{eqnarray}
  P_{,y} &=& -2 \frac{Q}{y^5}, \\[10pt]
  X_{,y} &=& -2 \frac{R}{y}, \\[10pt]
  Q_{,y} &=& -2 \alpha \frac{X^2P}{y}, \\[10pt]
  \label{minkRy}
  R_{,y} &=& -2X \left( \frac{P^2}{y} + 4\frac{X^2-1}{y^5} \right).
\end{eqnarray}
Standard second order
centered finite differencing results in $4(N-2)$ non-linear algebraic
equations which are supplemented by the 4 boundary conditions 
(\ref{mink_bound}) and 4 interface relations
\begin{eqnarray}
  P_{n_1+1} &=& P_{n_1}, \label{int_P} \\
  X_{n_1+1} &=& X_{n_1}, \\
  Q_{n_1+1} &=& Q_{n_1}, \\
  R_{n_1+1} &=& R_{n_1} \label{int_R}. 
\end{eqnarray}  
We then start with piecewise linear initial guesses for $P$ and $X$ (and the
corresponding derivatives $Q$ and $R$) and solve the $4N$ algebraic equations
iteratively with a Newton--Raphson method. Results
for various choices of the string parameter $\alpha$
are shown in Figure 6 of paper I.

\subsection{The static cosmic string coupled to gravity}

From the numerical point of view, the problem of solving for a static cosmic
string coupled to gravity  through the Einstein equations is virtually
identical to that of a static string in Minkowski spacetime.
The only difference is the much higher degree of complexity
of the equations due to the appearance of the functions 
$\nu$, $\tau$, $\mu$ and $\gamma$
as extra variables. We do not present the equations here, since 
they may be derived from the fully dynamic case by setting all time derivatives
to zero in the relevant equations. The solution is again obtained using the 
relaxation method described
in the previous section. As our initial guess for the metric
variables we use Minkowskian values, and for the string variables $X$
and $P$ we use the previously calculated values for a Minkowskian string with
the same string parameters. The results are shown in Figure 7 of paper I.

\subsection{The dynamic code}

In the dynamic case all variables $\nu$, $\tau$, $\mu$, $\gamma$, $P$ and $X$
are functions of $u,r$ and we have to solve the system 
(\ref{nuur})--(\ref{xur}) of partial differential equations (PDEs). 
In order to
control the behavior of the solution at infinity, we need a
generalisation for PDEs of the relaxation scheme applied to ordinary 
differential equations (ODEs). In view of the characteristic feature of the 
relaxation scheme, namely the simultaneous calculation of new function values 
at all grid points, this generalisation leads directly to implicit 
evolution schemes as used for hyperbolic or parabolic PDEs.
Therefore, the dynamic code is based on the implicit,
second order in space and time Crank--Nicholson scheme (see \cite{relax}
for example). For each solution step
we consider two spatial slices of the 
grid-type of Figure \ref{grid}, labelled $n$ and
$n+1$. We then apply centered finite differencing according to
\begin{eqnarray}
  f &=& \frac{f^{n+1}_{k+1} + f^{n+1}_k + f^n_{k+1} + f^n_k}{4}, \\[10pt]
  f_{,r} &=& \frac{f^{n+1}_{k+1} - f^{n+1}_k + f^n_{k+1} - f^n_k}{2\Delta r},
        \\[10pt]
  f_{,u} &=& \frac{f^{n+1}_{k+1} + f^{n+1}_k - f^n_{k+1} - f^n_k}{2\Delta u},
\end{eqnarray}
where $f$ represents any of our variables. Assuming that all functions are
known on slice $n$ we arrive at a large set of algebraic equations for the
$f^{n+1}_i$, similar to the
static case, which needs to be supplemented by boundary conditions 
and interface 
relations analogous to (\ref{int_P})--(\ref{int_R}). Again the system of
algebraic equations is solved iteratively with the Newton--Raphson method.
The initial guess for the data on the new slice $n+1$ is 
taken from the previous slice and convergence is typically achieved
within three iterations. For this purpose we rewrite the
dynamic equations (\ref{nuur})--(\ref{xur}) as a first order system.
The equations for the variables $\nu$, $\tau$ and $X$ involve radial
derivatives which may be written in terms of the second order operator
$\frac{\partial}{\partial r}(r\frac{\partial}{\partial r})$ so we
introduce the corresponding variables $N=r\nu_{,r}$, $T=r\tau_{,r}$ and
$R=rX_{,r}$ [cf. paper I equations (110)--(115)].
The equation for $\mu$ on the other hand involves
$\frac{\partial}{\partial r}(r^2\frac{\partial}{\partial r})$, while
that for $P$ involves $\frac{\partial}{\partial
r}(r^{-1}\frac{\partial}{\partial r})$. We therefore introduce the
corresponding variables $M=r^2\mu_{,r}$ and $Q=r^{-1}P_{,r}$.
Finally the equation for $\gamma$ involves only one
$r$ derivative and may be written in first order form without having
to introduce any further quantities. In terms of these variables
(\ref{nuur})--(\ref{xur}) become:

\begin{eqnarray}
   \nu_{,r}  &=& \frac{N}{r}, \label{nur} \\[10pt]
   \tau_{,r} &=& \frac{T}{r}, \\[10pt]
   \mu_{,r}  &=& \frac{M}{r^2}, \\[10pt]
   P_{,r}    &=& rQ,          \\[10pt]
   X_{,r}    &=& \frac{R}{r}, \\[10pt]
   2N_{,u}   &=& N_{,r} + \frac{T^2-N^2}{r\nu} + \frac{NM}{r^2}
                + 2 \frac{\nu_{,u} N - \tau_{,u} T}{\nu} -\nu_{,u} 
                - \frac{\nu_{,u} M}{r} - N\mu_{,u} \nonumber \\[10pt]
             && + 8\pi \eta^2 \left[ 2e^{2(\gamma + \mu)} r(X^2-1)^2
                + \frac{1}{\alpha} e^{-2\mu} \nu^2(2 P_{,u}Q - rQ^2)\right],
                \\[10pt]
   2T_{,u}   &=& T_{,r} - 2\frac{TN}{r \nu} + 2\frac{\tau_{,u} N+\nu_{,u}T}{\nu}
                + \frac{TM}{r^2} - \tau_{,u} - \frac{\tau_{,u} M}{r} 
                -T\mu_{,u}, \\[10pt]
   2M_{,u}   &=& M_{,r} + \frac{M^2}{r^2} - 2\mu_{,u} M - 2 r \mu_{,u}
                + 8\pi\eta^2 \left[ e^{2\gamma} X^2P^2 + 2 
                \frac{e^{2(\gamma+\mu)}}{\nu}r^2(X^2-1)^2 \right], \\[10pt]
   2(r+M)\gamma_{,r} &=& M_{,r} - 2\frac{M}{r} - \frac{M^2}{r^2}
                + \frac{T^2+N^2}{2\nu^2} + 8\pi \eta^2 \left[ R^2
                + \frac{1}{\alpha} e^{-2\mu}\nu r^2Q^2 \right],
                \label{1gammar} \\[10pt]
   2Q_{,u}   &=& Q_{,r} - \frac{QM}{r^2} + Q \mu_{,u} - \frac{Q\nu_{,u}}{\nu}
                - \frac{P_{,u} N}{r^2\nu} + \frac{QN}{r\nu} + \frac{P_{,u}}{r^2}
                + \frac{P_{,u} M}{r^3}-\alpha \frac{e^{2(\gamma+\mu)}}{\nu}
                \frac{PX^2}{r}, \\[10pt]
   2R_{,u}   &=& R_{,r} - X_{,u} - \frac{X_{,u} M}{r} + \frac{RM}{r^2} 
                - R\mu_{,u} - 4 \frac{e^{2(\gamma+\mu)}}{\nu} r X(X^2-1)
                - e^{2\gamma} \frac{XP^2}{r} \label{Rur}.
\end{eqnarray}
The corresponding first order system in the outer region is given by
\begin{eqnarray}
  \nu_{,y}  &=& -2 \frac{N}{y}, \label{nuy}  \\[10pt]
  \tau_{,y} &=& -2 \frac{T}{y}, \\[10pt]
  \mu_{,y}  &=& -2 yM,          \\[10pt]
  P_{,y}    &=& -2 \frac{Q}{y^5}, \\[10pt]
  X_{,y}    &=& -2 \frac{R}{y}, \\[10pt]
  2N_{,u}   &=& -\frac{1}{2}y^3 \left( N_{,y} -2yNM -2\frac{T^2-N^2}{y\nu}
               \right) -y^2 \nu_{,u} M - N\mu_{,u}
               +2\frac{\nu_{,u} N - \tau_{,u} T}{\nu} - \nu_{,u} 
               \nonumber \\[10pt]
            && \qquad \,\,\,  +8\pi \eta^2 \left[ 2e^{2(\gamma + \mu)}
               \frac{(X^2-1)^2}{y^2} + \frac{1}{\alpha} e^{-2\mu} \nu^2
               \left( 2P_{,u} Q - \frac{Q^2}{y^2}  \right) \right], \\[10pt]
  2T_{,u}   &=& -\frac{1}{2}y^3 \left( T_{,y} -2y TM+4\frac{TN}{y\nu}\right)
               - T\mu_{,u} -y^2 \tau_{,u} M
               + 2\frac{\tau_{,u} N + \nu_{,u} T}{\nu} - \tau_{,u}, \\[10pt]
  2M_{,u}   &=& -\frac{1}{2} y^3 (M_{,y} -2yM^2) - 2\frac{\mu_{,u}}{y^2}
               -2\mu_{,u} M + 8\pi\eta^2 \left[ e^{2\gamma} X^2P^2
               + 2 \frac{e^{2(\gamma+\mu)}}{\nu}
               \frac{(X^2-1)^2}{y^4} \right], \\[10pt]
  2(y^2M+1) \gamma_{,y} &=& \,\,\, y^2M_{,y} + 4yM +2y^3M^2 
               - \frac{N^2+T^2}{y\nu^2}
               -16\pi\eta^2 \left( \frac{R^2}{y} + \frac{1}{\alpha} e^{-2\mu}
               \nu \frac{Q^2}{y^5} \right), \\[10pt]
  2Q_{,u}   &=& -\frac{1}{2}y^3 \left( Q_{,y} +2yQM -2\frac{QN}{y\nu} \right)
               +y^4P_{,u} - y^4\frac{P_{,u} N}{\nu} - \frac{\nu_{,u} Q}{\nu}
               \nonumber \\[10pt]
            && +y^6 P_{,u} M + Q\mu_{,u}
               -\alpha \frac{e^{2(\gamma + \mu)}}{\nu} y^2PX^2, \\[10pt]
  2R_{,u}   &=& -\frac{1}{2}y^3 (R_{,y} -2yRM) -X_{,u} -R\mu_{,u}
               - y^2X_{,u} M - e^{2\gamma} y^2 XP^2
               - 4\frac{e^{2(\gamma+\mu)}}{\nu} \frac{X(X^2-1)}{y^2}.
               \label{Ruy}
\end{eqnarray}
In order to solve these equations we must supplement them by
appropriate initial and boundary conditions. We start by considering
boundary conditions on the axis. In general we find the code is more
stable if one imposes boundary conditions on the radial derivatives
rather than the variables themselves. For the variables $\nu$, $\tau$
and $X$ we therefore impose the required
boundary conditions on the initial data, but in the subsequent
evolution we impose the weaker condition that their radial
derivatives are finite on the axis. This ensures that the evolution
equations propagate the axial conditions given on the initial
data. For the variable $\mu$ we impose the condition that $M$ is zero
on the axis which is equivalent to the rather weak condition that
$r^2\mu_r$  vanishes there. The inverse power of $r$ in the
definition of $Q$ makes it unsuitable to specify the value of this
quantity at $r=0$ so in this case we work with the variable directly
and require that $P=1$ on the axis. 
Finally the variable $\gamma$ is given by a purely radial equation, so in
this case we must specify the value on the axis which is chosen to be zero
to ensure elementary flatness. Therefore at $r=0$ we require 
\begin{eqnarray}
  N &=& 0, \\
  T &=& 0, \\
  M &=& 0, \\
  \gamma &=& 0, \\
  P &=& 1, \\
  R &=& 0.
\end{eqnarray}
For the boundary conditions at null infinity we know that regular
solutions of the cylindrical wave equation have radial derivatives that decay 
faster than $1/r$ so that
we may take the variables $N$, $T$ and $R$, which satisfy a wave
type equation, to vanish at $y=0$. The asymptotics of $\mu$ are
slightly different due to the additional power of $r$ in the radial
derivative (similar to the spherically symmetric wave equation) but for a
regular solution $\mu_{,y}$ vanishes at null infinity. 
The $P$ equation does not satisfy a wave type equation due to the
inverse power of $r$ but has
asymptotic behavior given by a modified Bessel function. The
physically relevant finite solution has exponential decay so in this
case one may impose the condition that $Q=0$ at $y=0$. Hence we
require  the solution to satisfy the following boundary conditions at $y=0$ 
\begin{eqnarray}
  N &=& 0, \\
  T &=& 0, \\
  \mu_{,y} &=& 0, \\
  Q &=& 0, \\
  R &=& 0.
\end{eqnarray}
These boundary conditions are sufficient to determine the solution of
the first order system (\ref{nur})--(\ref{Ruy}) while suppressing the unphysical
solutions which are singular on the axis or null infinity. Note that
$\gamma$ is determined by the constraint equation
(\ref{gammar}), which is a first order ODE, and thus only needs one
boundary condition.

\section{Testing the code}
In paper I we used the CCM code to reproduce several exact vacuum 
solutions to an accuracy of about $10^{-5}$ with second order convergence. 
We have also checked the properties of the codes for the static cosmic string
in Minkowski spacetime and coupled to gravity.
Both clearly showed second order convergence. Here we will focus on testing
the implicit dynamic code. We have carried out four different independent 
tests, namely 

\newcounter{count}
\begin{list}{\rm{(\arabic{count})}}{\usecounter{count}
             \labelwidth1cm \leftmargin1.5cm \labelsep0.4cm \rightmargin1cm
             \parsep0.5ex plus0.2ex minus0.1ex \itemsep0ex plus0.2ex}
\item Reproducing the non-rotating vacuum solution of Weber and Wheeler
\cite{WW},
\item Reproducing the rotating vacuum solution of Xanthopoulos \cite{Xan},
\item Using the results for the static cosmic string (paper I)
      as initial data and checking that the system stays in its static
      configuration,
\item Convergence analysis for the string hit by a Weber--Wheeler wave.
\end{list}

Two additional tests arise in a natural way from the field equations and
the numerical scheme. As described above there are two additional
field equations which are algebraic consequences of the other field equations.
We have verified that these equations are satisfied to second order accuracy
($\sim \Delta r^2$). Furthermore the numerical scheme calculates the residuals
of the algebraic equations to be solved, which have thus been monitored
in test runs. They are satisfied to a much higher accuracy 
(double precision
machine accuracy), so the total error is dominated by the truncation error
of the second order differencing scheme. Another independent test is the
comparison with the explicit CCM code which yields good
agreement for as long as the latter remains stable.
The four main tests are now described in more detail.

\subsection{The Weber--Wheeler wave}

In the first test we evolve the analytic solution given by Weber and Wheeler 
\cite{WW}, which describes a gravitational pulse of the + polarisation mode.
This solution has two free parameters, $a$ and $b$, which can be interpreted as
the width and amplitude of the pulse. The equations together with
a more detailed discussion have been given in paper I.
%
%
%
%
%
\begin{figure}[t]
  \centering
  \epsfig{file=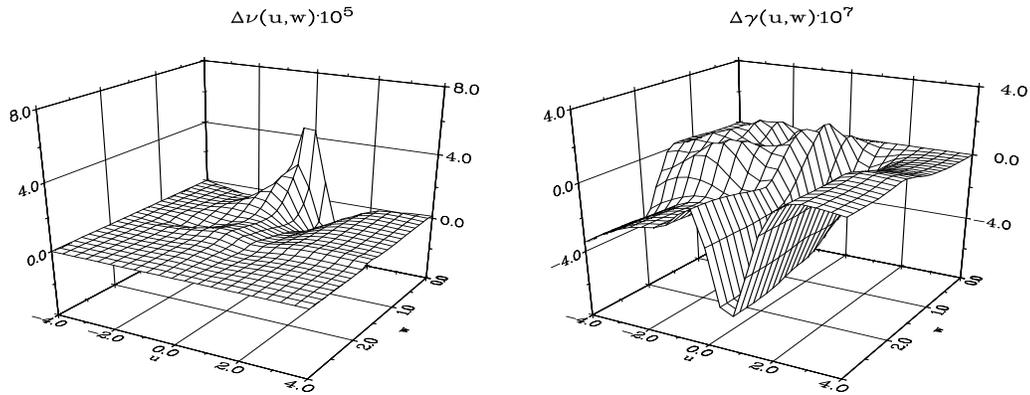, height=400pt, width=150pt, angle=-90}
  \caption{The deviation of the numerical $\nu$ and $\gamma$ from the
           Weber--Wheeler solution as a function of $u$ and $w$ 
           obtained for 1920 grid points ($n_1=320$,
           $n_2 = 1600$). The wave parameters are $a=2$, $b=0.5$.
           Note that the error is amplified by $10^5$ and $10^7$ respectively.}
\label{plot_ww_delta}
\end{figure}
%
We prescribe $\nu$ as initial data according
to the analytic expressions obtained for $a=2$ and $b=0.5$ and set the other
free variables to zero, while $\gamma$ is calculated
via quadrature from the constraint equation (\ref{gammar}).
In Figure \ref{plot_ww_delta} we show
the deviation of the numerical results from the analytical one for
$N=1920$ grid points (320 points in the inner, 1600
points in the outer region) and a Courant factor of 0.5 with respect to
the inner region. The convergence analysis (see below)
shows that this number of points provides sufficient resolution in the outer
region while still keeping computation times at a tolerable level.
All computations presented in this work have been obtained
with these grid parameters, unless stated otherwise.
The code stays stable for much longer time intervals than shown in the figure,
but does not reveal any further interesting features as the analytic solution
approaches its Minkowskian values and the error goes to zero.

\subsection{The rotating solution of Xanthopoulos}

Xanthopoulos \cite{Xan} derived an analytic vacuum solution for Einstein's
field equations in cylindrical symmetry containing both the + and $\times$
polarisation mode. Its analytic form in terms of
our metric variables and a more detailed discussion has been given in paper I.
The solution has one free parameter $a$ which is set to one in this 
calculation.
%
%
\begin{figure}[t]
  \centering
  \epsfig{file=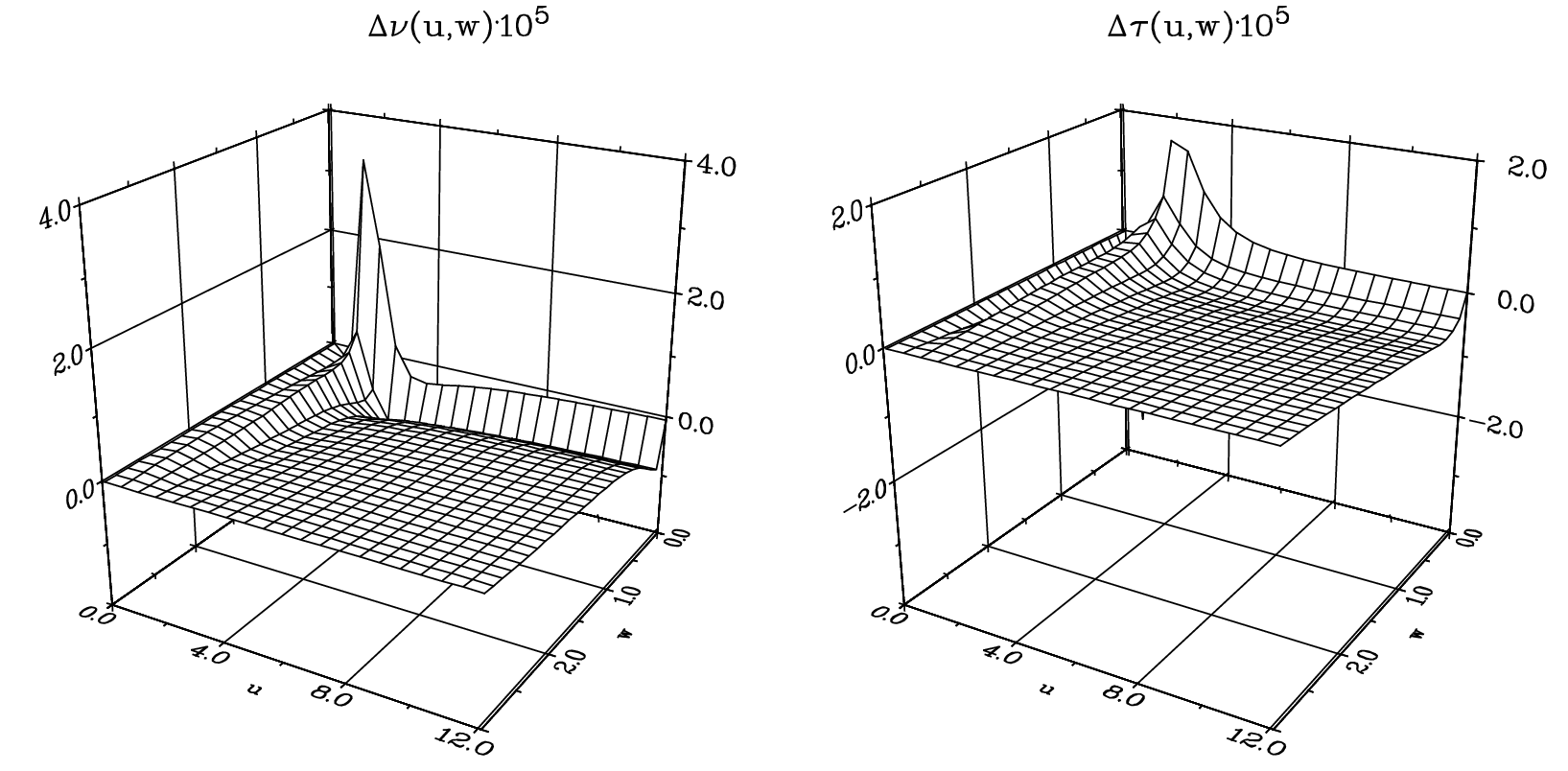, height=400pt, width=150pt, angle=-90}
  \epsfig{file=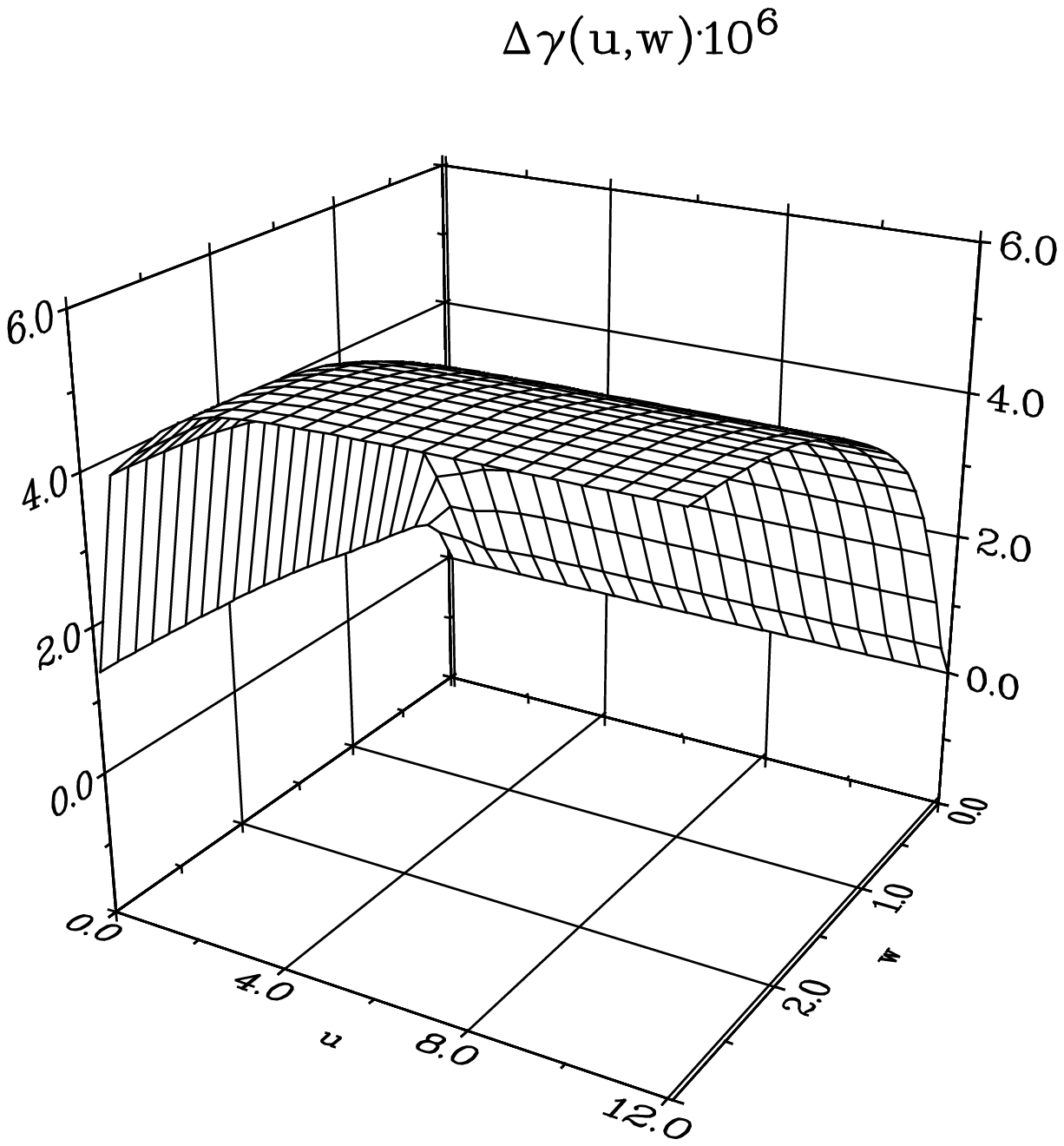, height=200pt, width=150pt, angle=-90}
  \caption{The deviation of the numerical $\nu$, $\tau$, $\gamma$ from
           Xanthopoulos' analytic solution as a function of $u$ and $w$
           obtained for 1920 grid points
           ($n_1=320$, $n_2 = 1600$).
           Note that the error is amplified by $10^5$ and $10^6$ 
           respectively.}
\label{plot_as_delta}
\end{figure}
\begin{figure}[t]
  \centering
  \epsfig{file=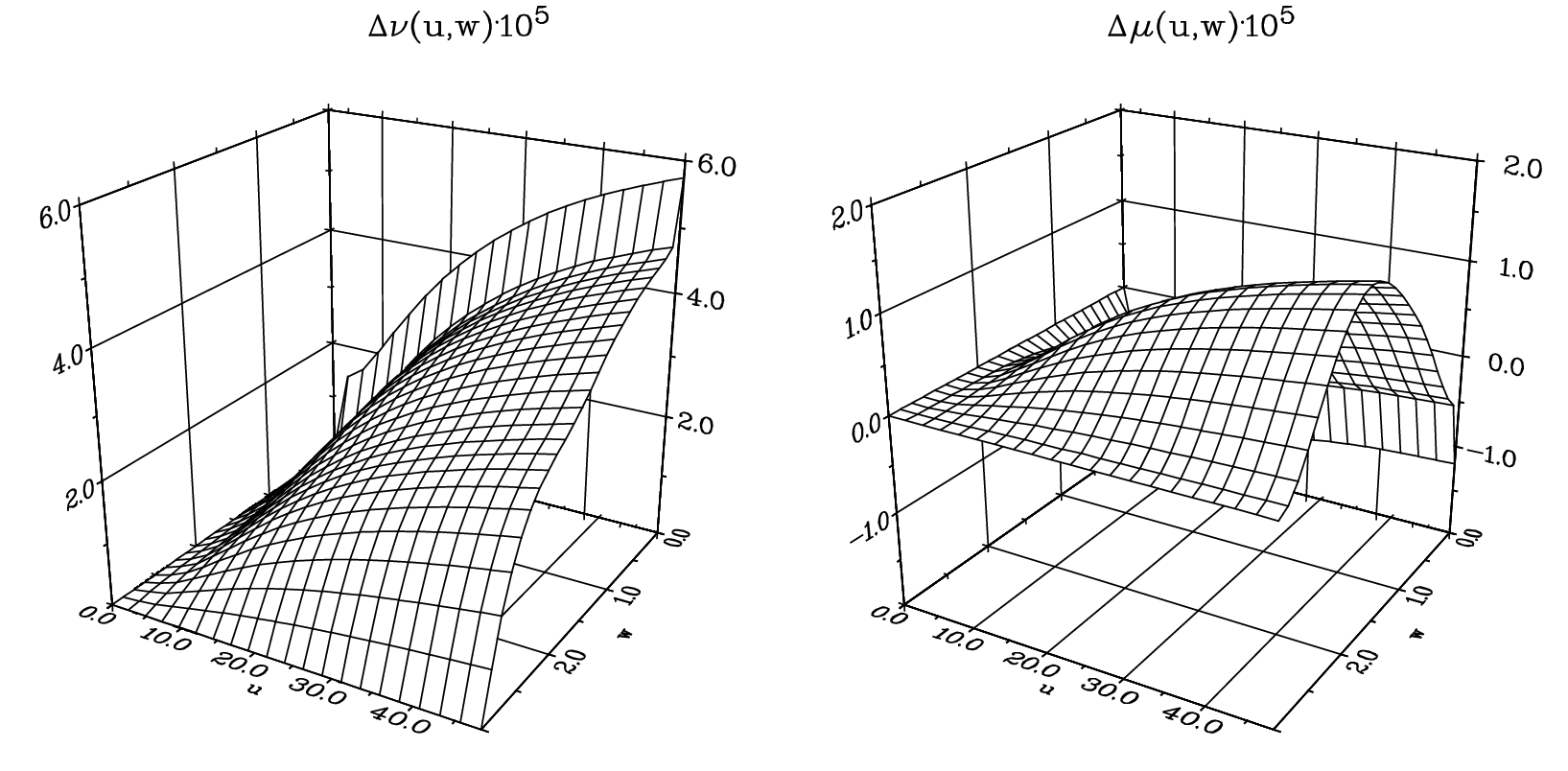, height=400pt, width=150pt, angle=-90}
  \epsfig{file=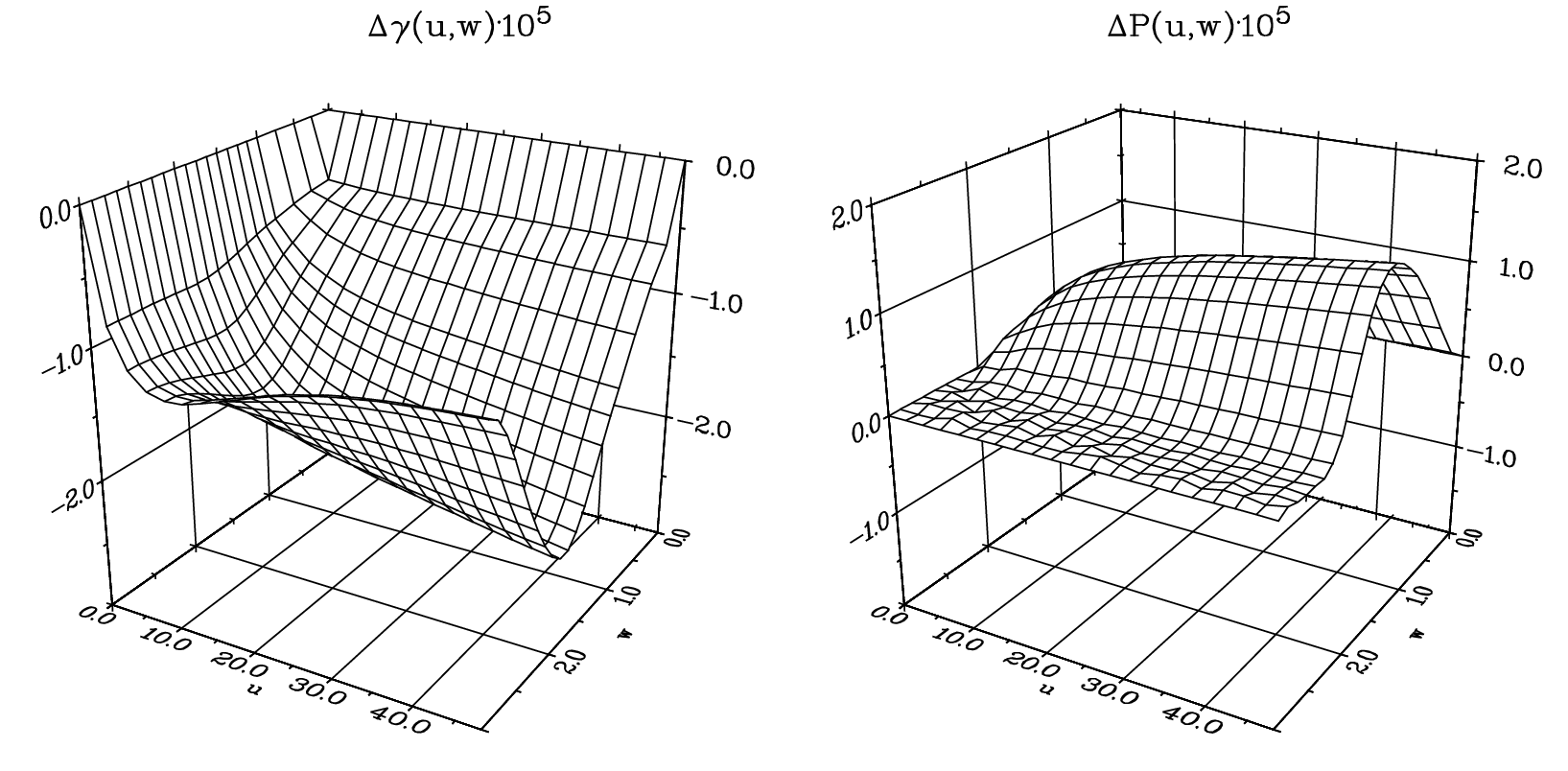, height=400pt, width=150pt, angle=-90}
  \epsfig{file=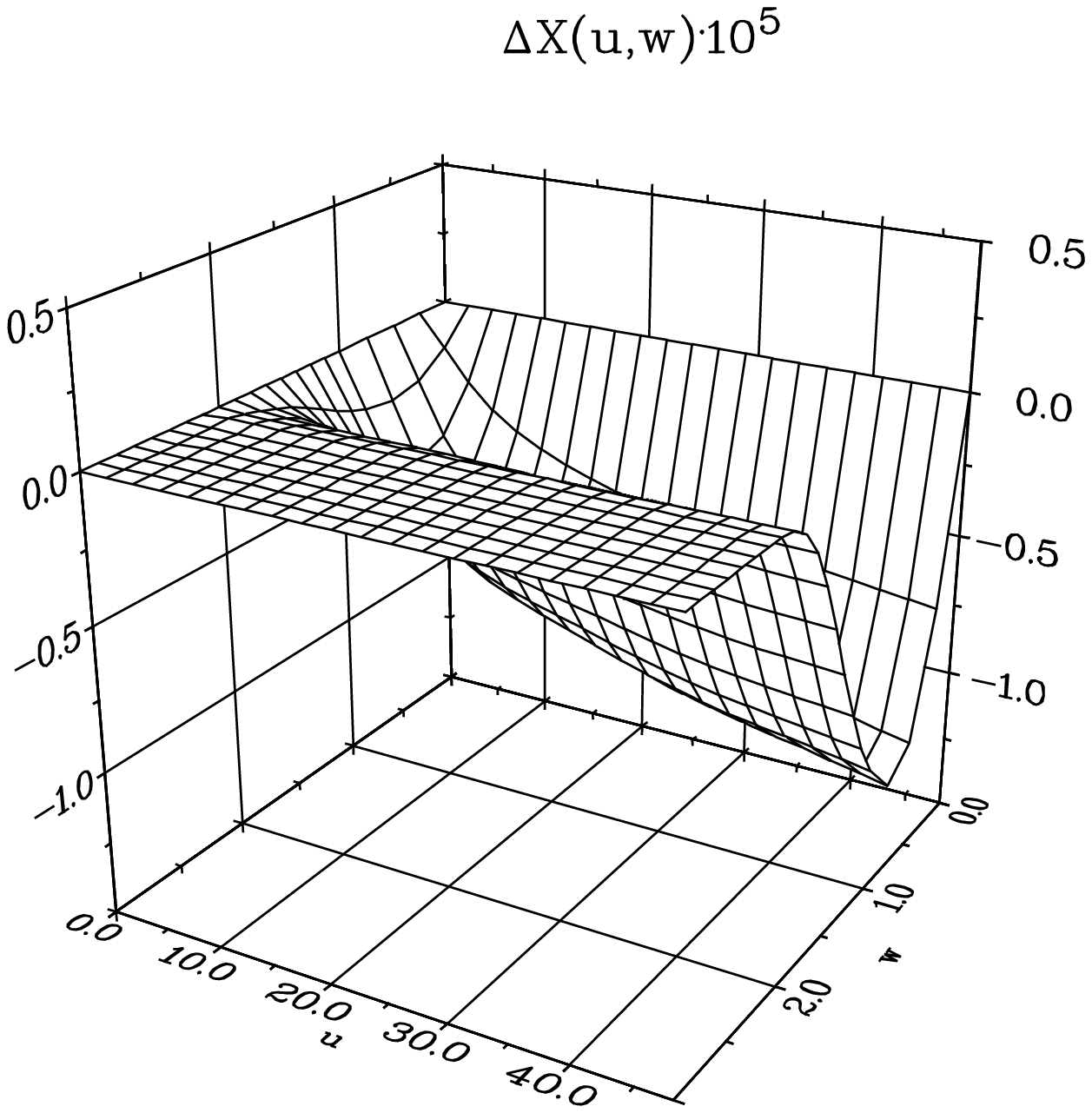, height=200pt, width=150pt, angle=-90}
  \caption{The deviation of the metric and matter variables from the initial
           data in the case of evolving a static cosmic string with
           $\alpha=1$, $\eta=0.2$. For our standard grid with 
           $1920$ points, the
           configuration stays static to an accuracy of about $10^{-5}$                  over a range of more than 30000 time steps.}
  \label{evolstat}
\end{figure}
%
%
%
%
%
The error of our numerical results is displayed in Figure \ref{plot_as_delta},
where we have used the same grid parameters as in the Weber--Wheeler case.
Again we have run
the code for longer times and found that the error approaches zero.
We conclude that the code reproduces both analytic vacuum solutions
with excellent accuracy comparable to that of the CCM code
and exhibits long term stability.

\subsection{Evolution of the static cosmic string}

\begin{figure}[t]
  \centering
  \epsfig{file=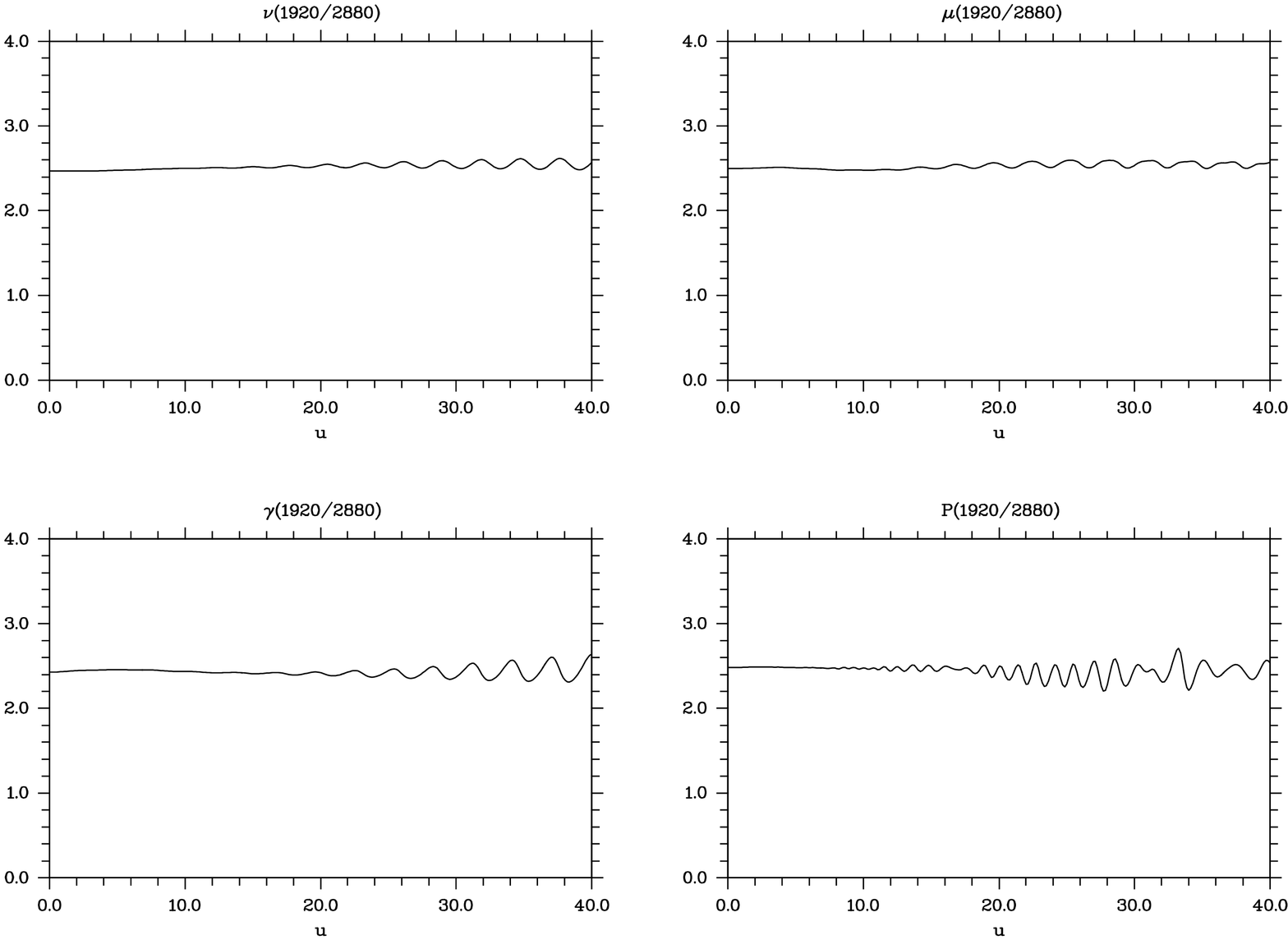, height=400pt, width=283pt, angle=-90}
  \epsfig{file=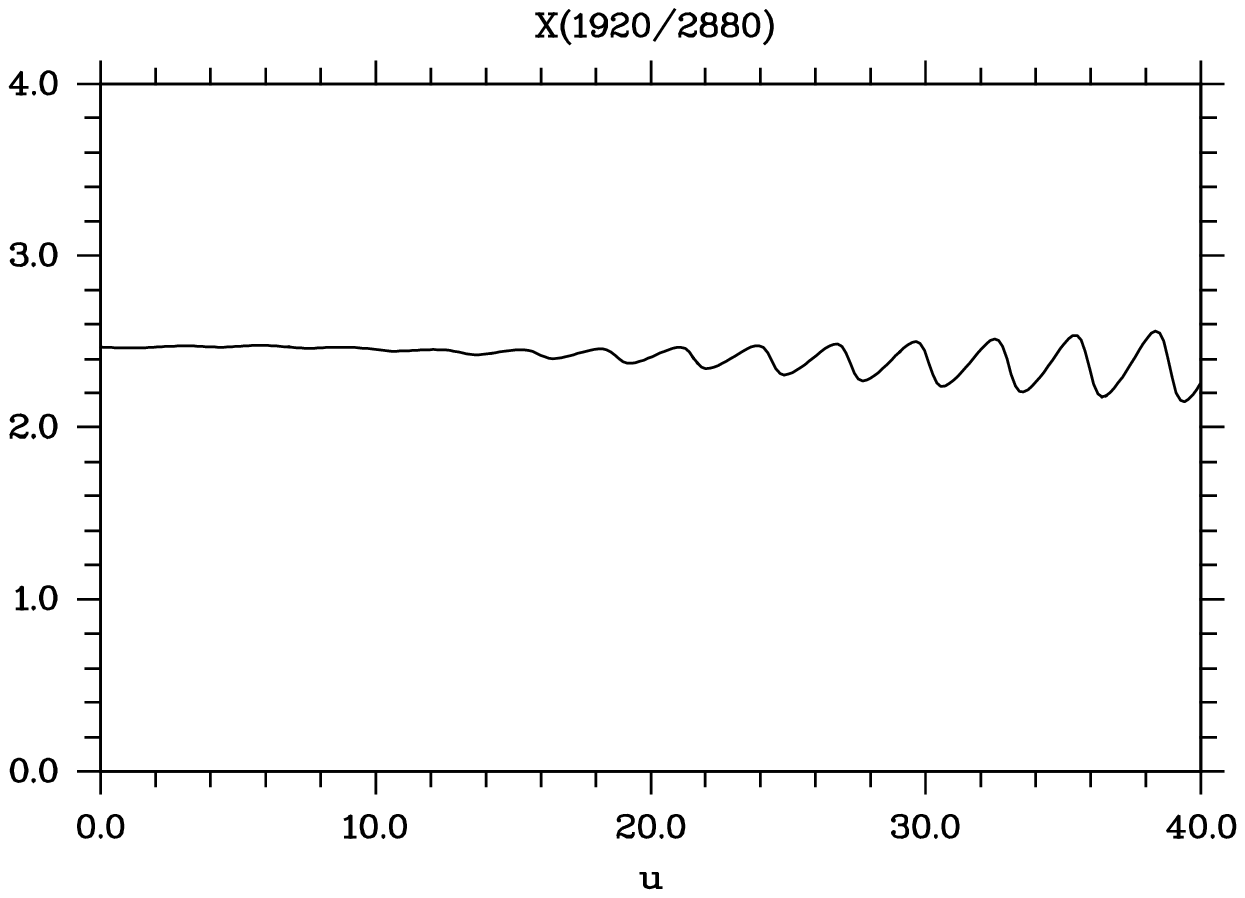, height=200pt, width=141pt, angle=-90}
  \caption{The convergence factor $\ell_2[\Psi^{1920}] / \ell_2[\Psi^{2880}]$
           is plotted as a function of $u$. 
           We expect a convergence factor of 2.25 since the number of 
           grid points is multiplied by 1.5.
           Even though our results show weak variability at later times, 
           second order convergence is maintained
           throughout long runs (more than 30000 time steps with $N=1920$).}
\label{cs_conv}
\end{figure}

The tests described above only involve vacuum solutions, so the matter 
part of the
code and the interaction between matter and geometry has not been tested.
An obvious test involving matter and geometry is to use the result for the
static cosmic string in curved spacetime as initial
data and evolve this scenario. All variables should, of course, remain
at their initial values. We have evolved the static string data for
our standard grid and the parameter set, $\alpha = 1$ and $\eta=0.2$,
which corresponds to a strong back-reaction of the string on the
metric. The results are shown in Figure \ref{evolstat}. The system stays
in its static configuration with high accuracy over a long time interval.

\subsection{Convergence analysis}
Our investigation of the interaction between the cosmic string and gravitational
waves will focus on the string being hit by a wave of the Weber--Wheeler type.
In order to check this scenario for convergence we have run
the code for the parameter set $\eta=0.2$, $\alpha=1$, $a=2$, $b=0.5$
for different grid resolutions, where $a$ and $b$ 
are again the width and amplitude
of the Weber--Wheeler wave. In our case it is of particular interest to
investigate the time dependence of the convergence to see what resolution we
need in order to obtain reliable results for long runs.
We calculate the convergence rate in the same way as in the static case
(cf. paper I), but this time the $\ell_2$-norm is
a function of time. So for each variable we have
\begin{eqnarray}
  \ell_2[\Delta \Psi^N](u) &=& \sqrt{\frac{\sum [\Psi^N_k(u) 
                               -\Psi^{4320}_k(u)]^2} {N}}, \label{l2(u)}
\end{eqnarray}
where the upper label ``4320'' indicates that the high resolution reference
solution has been calculated for $N=4320$ grid points.
In Figure \ref{cs_conv} we show the convergence factor
$\ell_2[\Psi^{1920}]/\ell_2[\Psi^{2880}]$ as a function of $u$
for $\nu$, $\mu$, $\gamma$, $P$ and $X$. The initial data for $\tau$ is 
identically zero for this scenario and stays zero during the evolution.
The number of grid points is increased by a factor of 1.5 here (instead of the
more commonly used 2) to reduce the computation time.
Only points common to all grids have been used in the
sum in equation (\ref{l2(u)}). For second order convergence we would expect
a convergence factor of $1.5^2$. Although the results in
Figure \ref{cs_conv}
show weak variations with $u$, second order convergence is clearly maintained
for long runs.
In each case the outer region contains 5 times as many grid points as
the inner region (e.g. $n_1=320$, $n_2=1600$ for the $N=1920$ case). The
reason for this is that in the dynamical evolutions
$X$ and especially $P$ exhibit significant spatial variations
out to large radii. Due to the compactification, the spatial resolution of
our grid decreases as we move towards null infinity and to resolve the spatial
variations of the string variables out to sufficiently large radii we
therefore have
to introduce a large number of grid points in the outer region.
No such problems occur in the inner region. If significantly
fewer grid points are used in the outer region for this analysis, the
convergence properties of the string variables can deteriorate
to roughly first order level.

\section{Time dependence of the string variables}
\subsection{Static cosmic strings excited by gravitational waves}

\begin{figure}[t]
  \centering
  \epsfig{file=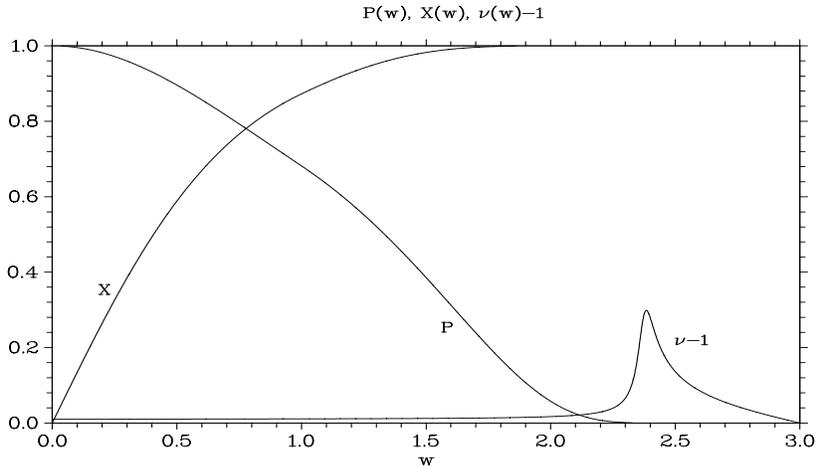, height=350pt, width=200pt, angle=-90}
  \caption{The initial data for $\nu$, $P$ and $X$ at $u_0=-20$
           for the standard parameters $\alpha=1$, $\eta=10^{-3}$,
           $a=2$, $b=0.5$. 
           The gravitational wave pulse is located in a region
           where the string fields $P$ and $X$ have almost fallen off
           to their asymptotic values.} 
\label{ini_nps0001}
\end{figure}
\begin{figure}[t]
  \centering
  \epsfig{file=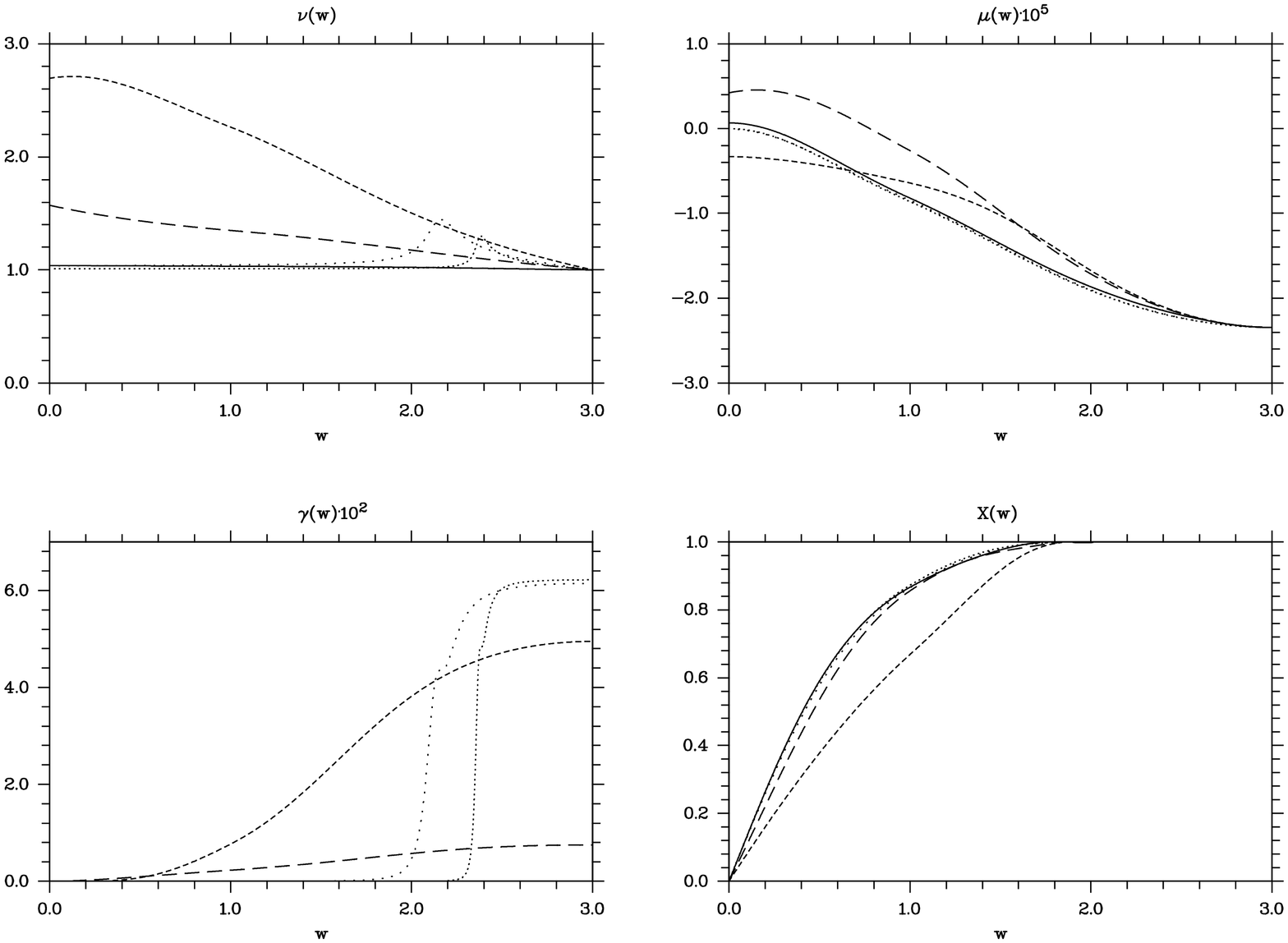, height=450pt, width=300pt, angle=-90}
  \epsfig{file=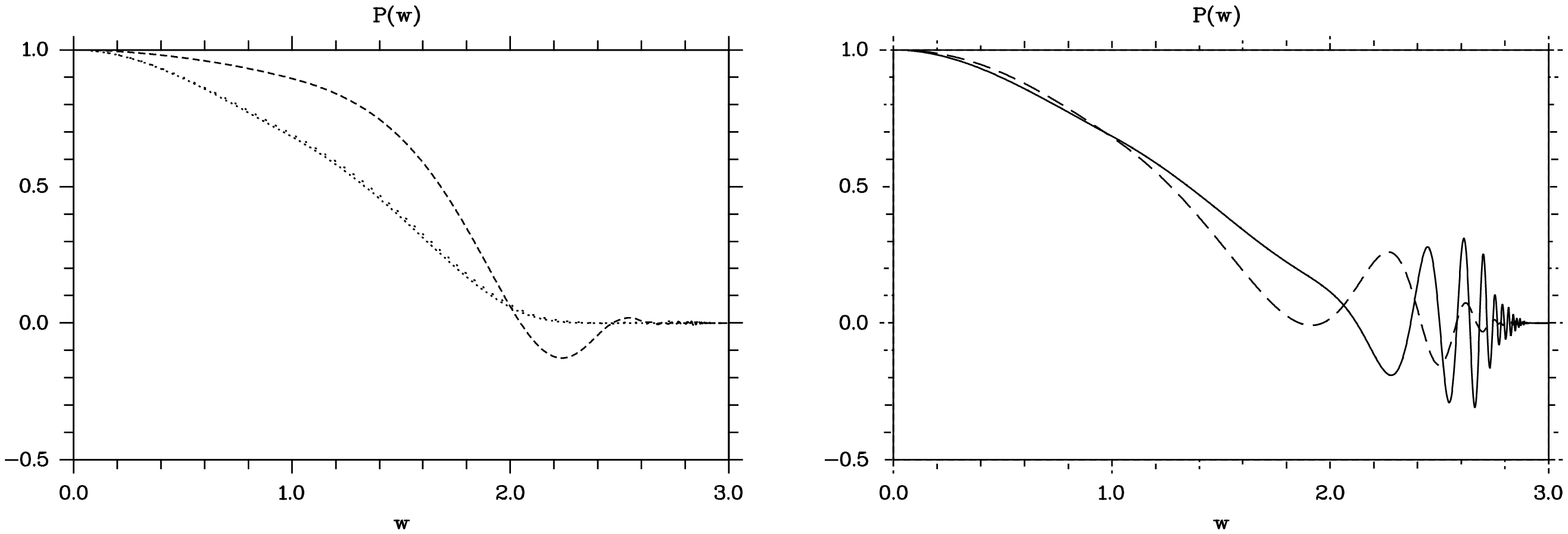, height=450pt, width=150pt, angle=-90}
  \caption{The metric and string variables are plotted as functions
           of $w$ for $u=-20$ (dotted), $u=-10$ (long dotted),
           $u=0$ (dashed), $u=2$ (long dashed) and $u=10$ (solid line).
           For clarity the the graphs of $P$ are distributed over two panels.
           The wave pulse (in $\nu$) initially moves inwards. It excites
           the string, is reflected at the origin and travels outwards.
           After $u=10$ only $P$ differs significantly from the
           static configuration as the oscillations
           slowly decay and propagate towards larger radii (cf. Figure
           \ref{ringing}).}
\label{mplot}
\end{figure}
\begin{figure}[t]
  \centering
  \epsfig{file=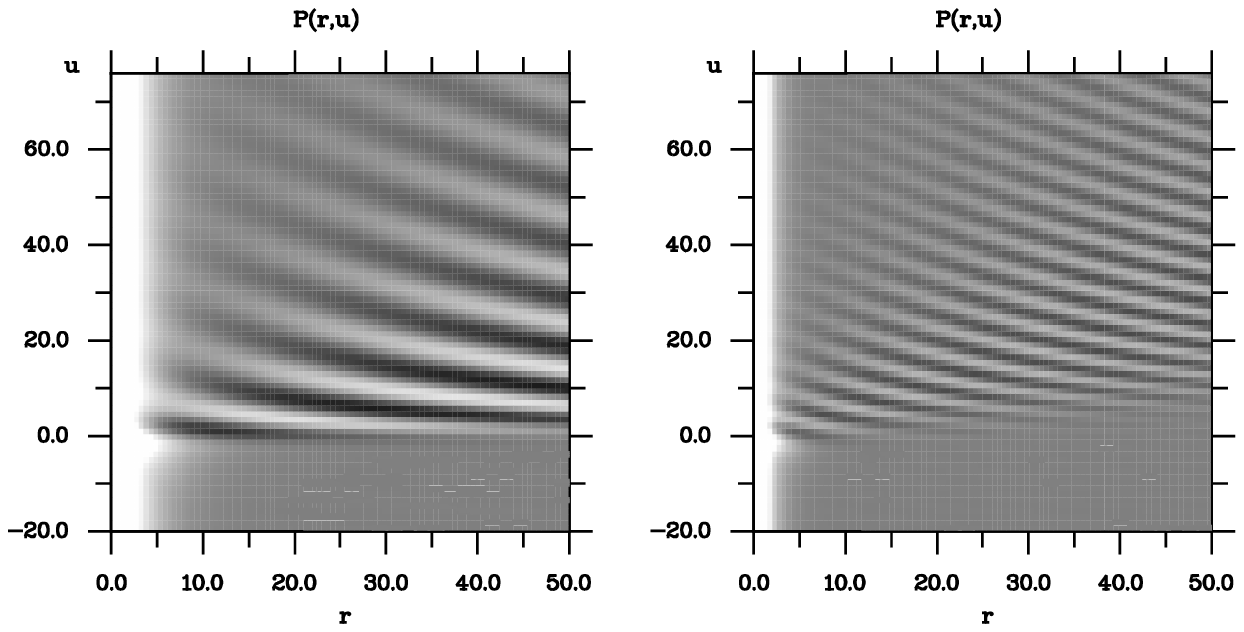, height=450pt, width=200pt, angle=-90}
  \caption{The cosmic string variable $P$ is shown as a function
           of radius and time for $\alpha = 0.2$ (left) and
           $\alpha = 1$ (right) (all other parameters have standard
           values). Note that we use the radial
           variable $r$ out to $r=50$ here. The ringing can clearly
           be seen and shows a lower frequency for smaller $\alpha$.}
\label{ringing}
\end{figure}
The scenario we are going to investigate in this section is an initially
static cosmic string hit by a gravitational wave of
Weber--Wheeler type. For this purpose we use the static results with
two modifications as initial data.
Firstly the static metric function $\nu_0$ is multiplied
by the exact Weber--Wheeler solution to simulate the gravitational wave
pulse. Thus we guarantee that initially the cosmic string is indeed
in an equilibrium configuration provided the wave pulse is located
sufficiently far away from the origin and its interaction with the string
is negligible.
Ideally the numerical calculation would start with the incoming wave located
at past null infinity. In order to approximate this scenario, we
found it was sufficient to use the large negative initial time $u_0=-20$. 
The second modification is to calculate $\gamma$ from the constraint
equation (\ref{gammar}) to preserve consistency with the Einstein field
equations.
In Figure \ref{ini_nps0001}
the corresponding initial data for $\nu$, $P$ and $X$ are shown for 
the parameter set $\eta=10^{-3}$, $\alpha=1$, $a=2$ and $b=0.5$. 
From now on we will refer to these values as ``standard parameters''
and only specify parameters if they take on non-standard values.
Note that $\tau$ vanishes on the initial slice in this case and stays 
identically zero throughout the evolution. The case of rotating gravitational
waves hitting a cosmic string will be analysed in a future publication.
The time evolution of the ``standard configuration'' is shown in Figure
\ref{mplot} where we plot $\nu$, $\mu$, $\gamma$, $P$ and $X$
as functions of $w$ at times $-20$, $-10$, $0$, $2$ and $10$. While the wave
pulse is still far away from the origin, its interaction with the cosmic
string is negligible (dotted lines). 
When it reaches the core region, however, it excites
the cosmic string and the scalar and vector field start oscillating (dashed
curves).
After being reflected at the origin, the wave pulse travels along the
outgoing characteristics and the metric variables
$\nu$, $\mu$ and $\gamma$ quickly settle down into their static configuration
which is close to Minkowskian values for $\eta = 10^{-3}$.
The vector and scalar field of the cosmic string, on the other hand, 
continue ringing albeit with a different character. Whereas the 
oscillations of the scalar field $X$ are dominant in the range $r\le 2$
and have significantly decayed at $u=10$ as shown in the figure,
the vector field oscillations propagate to large radii and fall off very
slowly (solid curves). This behavior is also
illustrated in the right panel of Figure \ref{ringing} which shows
a contour plot of $P$ as a function of $(u,r)$ out to $r=50$.
We shall see below, that the oscillations of 
$P$ will also decay eventually and the
cosmic string will asymptotically settle back into its equilibrium
configuration.

\subsection{Frequency analysis}

\begin{figure}[t]
  \centering
  \epsfig{file=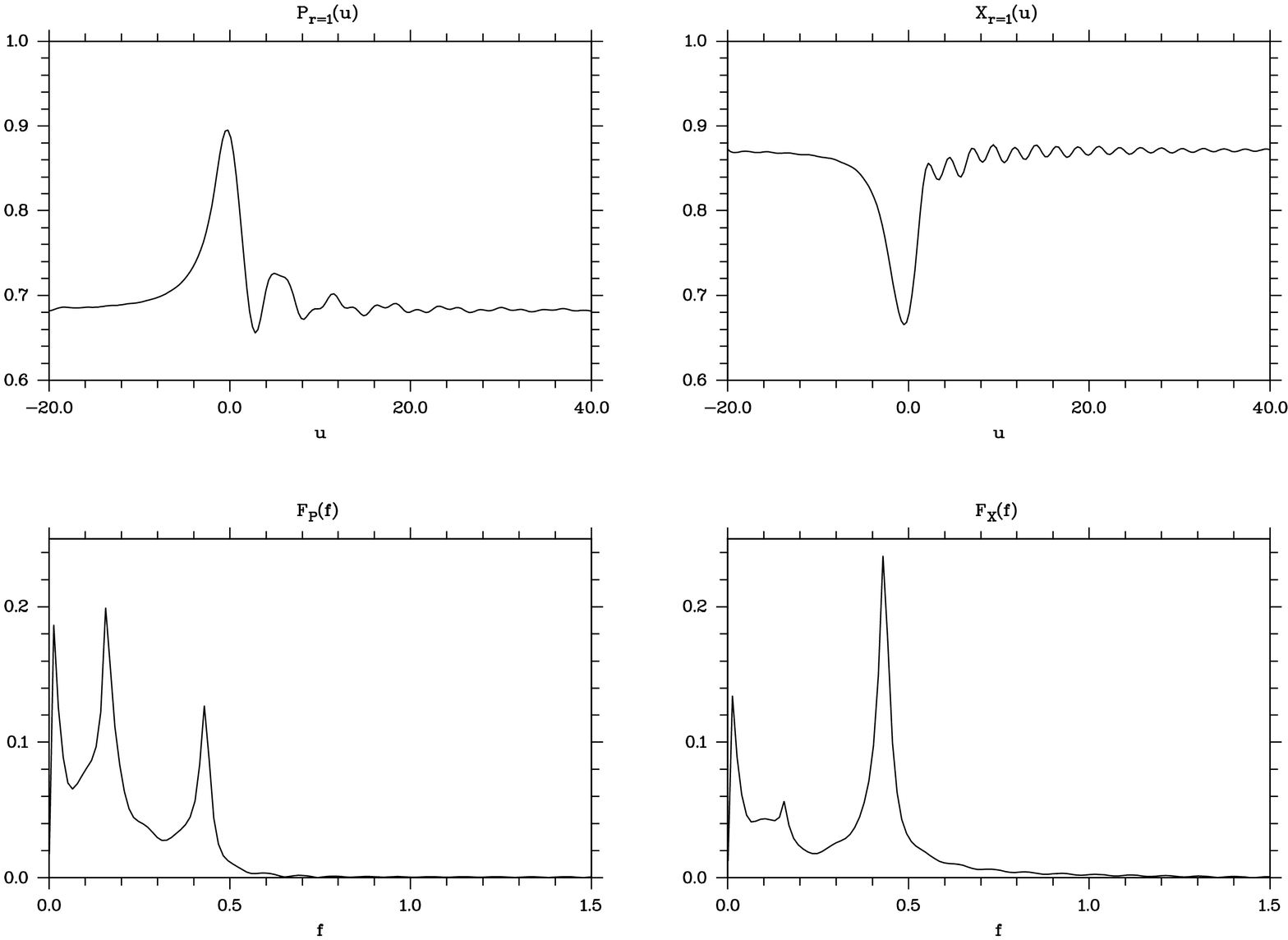, height=400pt, width=283pt, angle=-90}
  \caption{Upper panels: The variables $P$ and $X$ at $r=1$ are plotted
           as functions of $u$ for $\alpha=1$, $\eta=10^{-3}$, $a=2$ 
           and $b=0.5$. Lower panels: The corresponding power
           spectra. The spectrum for $P$ shows a strong peak at
           $f=0.16$, that for $X$ at $f=0.43$. The latter component, however,
           is also clearly present in the spectrum of $P$ which we attribute
           to the interaction between the fields (see text for details).
           Note that due to our rescaling of the coordinates, $u$ is measured
           in units of $1/\sqrt{\lambda} \eta$.}
\label{fourier}
\end{figure}

We will now quantitatively analyse the oscillations of the scalar and
vector field of the cosmic string. Since we are working in rescaled
coordinates, time and distance are
measured in units of $1/\sqrt{\lambda} \eta$ and 
frequency in its inverse. To avoid complicated notation, however,
we will omit the units from now on unless the meaning is unclear.
In order to measure frequencies, we Fourier analyse the time evolution
of the scalar and vector field for fixed radius $r$. Figure \ref{fourier}
shows $P$ and $X$ for standard parameters
as functions of time at $r=1$ together with the corresponding
power spectra. The plots for $X$ show a characteristic frequency $f_X=0.43$
whereas for $P$ we find a strong peak at $f_P=0.16$. The spectrum for $P$,
however, also shows a strong mode at $f=0.43$. We have calculated similar
power spectra for a large class of parameter sets and discovered this
effect on numerous occasions --- in addition to a strong peak at the
characteristic frequency of $P$ or $X$ there is a second maximum
at the characteristic frequency of the other field. In general the
characteristic mode of $X$ resulted in stronger peaks at smaller radius,
that of $P$ was stronger at larger radii.
We attribute this feature to the interaction between the scalar and vector
component of the cosmic string. The variation of the relative strength of
the oscillations with radius 
confirms the corresponding observation in Figure \ref{mplot}.
The accuracy of the measurements of the
frequencies is limited by the resolution of the Fourier spectra 
which again is limited by the time interval covered in the evolution and,
thus, by computation time. For tolerable computation times we get an
accuracy $\Delta f \approx 0.01$ which corresponds approximately to one bin
in the frequency spectra.

\begin{figure}[t]
  \centering
  \epsfig{file=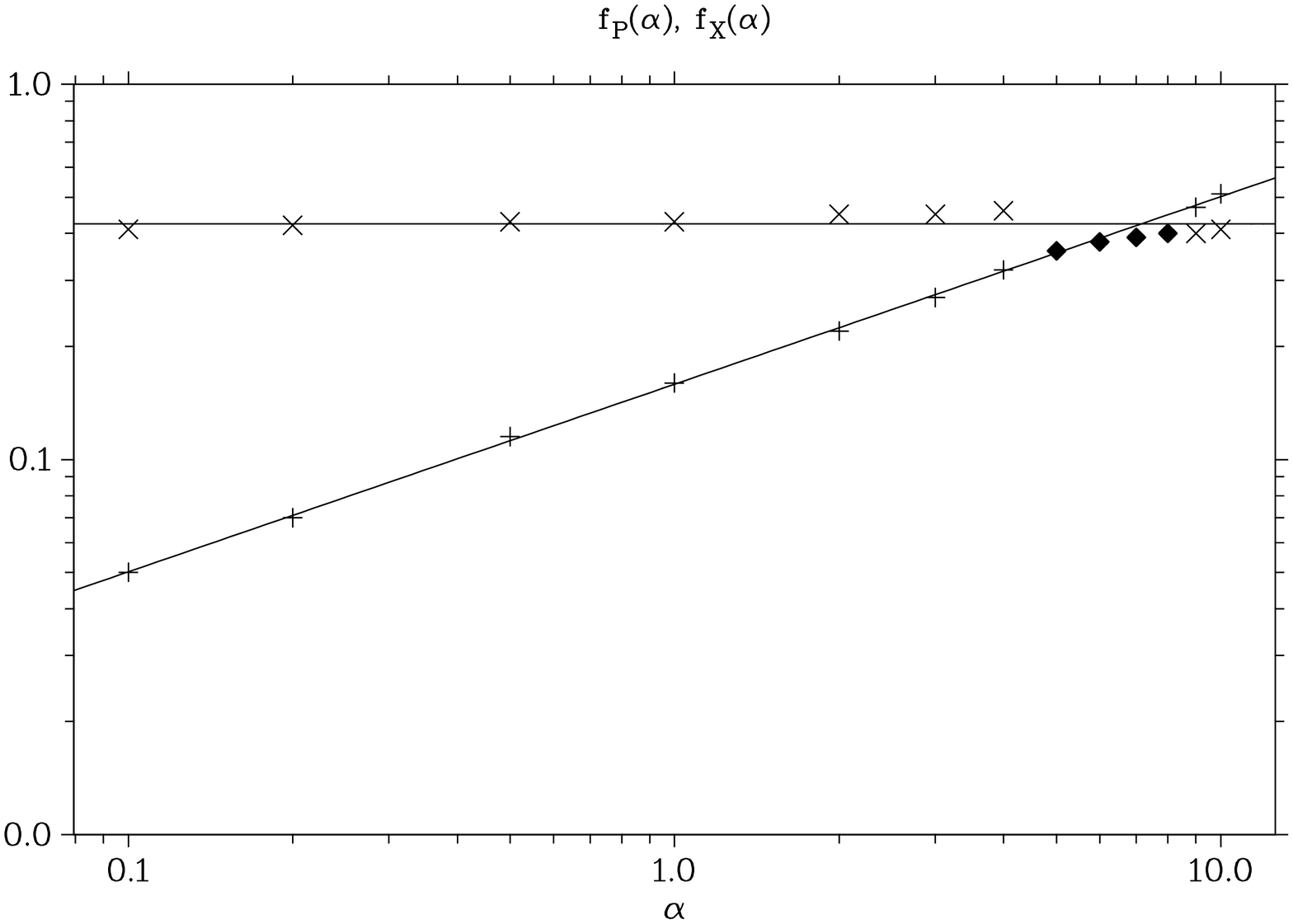, height=300pt, width=200pt, angle=-90}
  \caption{The frequencies $f_P$ ('+') and $f_X$ ('$\times$') are plotted as
           functions of $\alpha$. The curves have been obtained from
           fitting power laws. For $5 \le \alpha \le 8$ the expected
           frequencies for $P$ and $X$ are similar and we found only
           one maximum in the power spectra. These values are plotted
           as filled lozenges and have not been included in the regression 
           analysis as their classification is not obvious. The power
           law for $P$ coincides with the observed values to within
           measurement accuracy. The frequencies of $X$, however, show
           a significant, albeit small, deviation from the fitted
           constant line.}
\label{f_alpha}
\end{figure}

In order to investigate the dependency of the oscillations on 
$\alpha$, $\eta$, $a$, $b$ and the radial position $r$, we have varied each 
parameter over at least two orders of magnitude while keeping the other
parameters at standard values. We have found the following dependencies:
\begin{list}{\rm{(\arabic{count})}}{\usecounter{count}
             \labelwidth1cm \leftmargin1.5cm \labelsep0.4cm \rightmargin1cm
             \parsep0.5ex plus0.2ex minus0.1ex \itemsep0ex plus0.2ex}
\item The frequencies of both $X$ and $P$ did not show any variations
      with $\eta$ for $\eta < 0.1$. (Note that $\eta$ does, however,
      appear in the units). For larger values of $\eta$, the non-linear
      interaction between string and geometry becomes dominant and we did
      not find a simple relation between frequency maxima and parameters.
\item The variation of the parameters $a$ and $b$, 
      the width and amplitude of the Weber--Wheeler pulse, has 
      no measurable effect on the frequencies of $X$ and $P$, but
      only determined the amplitude of the oscillations. A narrow, strong pulse
      leads to larger amplitudes.
\item For small $r$ the oscillations in $X$ are stronger, whereas
      those for $P$ dominate at large $r$. The frequency values, however,
      do not depend on the radius. For radii greater than 10 
      the oscillations
      in $X$ had decayed so strongly that we could no longer measure its
      frequency.
\item To first order approximation the frequency
      of the scalar field, $f_X$, is independent of $\alpha$
      over the observed range $0.1 \le \alpha \le 10$. $f_P$, however,
      shows a strong dependence on $\alpha$ which is also illustrated
      in Figure \ref{ringing}, where we compare contour plots of $P$
      obtained for $\alpha=0.2$ and $1$. The frequency is significantly
      larger for $\alpha=1$.
\end{list}
In Figure \ref{f_alpha}
we show $f_P$ and $f_X$ as functions of $\alpha$ together with
the power laws obtained from a linear regression of the corresponding
double logarithmic data. In the range $5 \le \alpha \le 8$ (where
$\alpha=8$ corresponds to the critical coupling), the expected
values of $f_P$ and $f_X$ become similar and we observed only one maximum
in the corresponding power spectra. Therefore a classification with
respect to the vector or scalar origin of the frequencies is not obvious.
These values (shown by filled lozenges in Figure \ref{f_alpha}) 
have not been used
in the regression analysis. We obtain power law indices $\sigma_X = 0.00$
and $\sigma_P = 0.50$, so that
\begin{eqnarray}
  f_X & \sim & {\rm const.} \label{fX}\\
  f_P & \sim & \sqrt{\alpha}. \label{fP}
\end{eqnarray}
Whereas the fitted curve for $f_P$ coincides with the observed values
to high accuracy, Figure \ref{f_alpha} shows a small but significant
deviation of the measured $f_X$ from the fitted constant line. Indeed the
sizable range over $\alpha$ for which we observe only one frequency
indicates significant non-linear interaction similar to the effect of 
phase locking in non-linear systems of ODEs \cite{phase}.
If we consider $f_X$ and $f_P$ to be given by (\ref{fX}) and (\ref{fP}) in
terms of the rescaled unphysical coordinates and we transform this back
into  physical units using $\alpha=e^2/\lambda$, we arrive at the following 
relations for the physical variables
\begin{eqnarray}
  f_X & \sim & \sqrt{\lambda} \eta, \\[15pt]
  f_P & \sim & e \eta.
\end{eqnarray}
As shown in \cite{shellard} up to constant factors $\sqrt{\lambda} \eta$ and 
$e \eta$ are the masses
of the scalar and the vector field, $m_X$ and $m_P$, and we conclude that
$X$ and $P$ have characteristic frequencies
\begin{eqnarray}
  f_X & \sim & m_X, \\[15pt]
  f_P & \sim & m_P.
\end{eqnarray}
Since the frequencies for $X$ and $P$ seem only to depend upon the
respective masses we have attempted to confirm these results by
considering the oscillations of a cosmic string in two further scenarios.
Firstly since the frequencies do not depend upon the Weber--Wheeler
pulse we take as initial data the static values for the metric
variables but excite the string by adding a Gaussian perturbation to
either the $X$ or $P$ static initial values. The evolution is then
computed using the fully coupled system. Secondly since the
frequencies do not seem to depend upon the strength of the coupling to
the gravitational field we have completely decoupled the gravitational
field and considered the evolution of a cosmic string in Minkowski
spacetime. The initial data is taken to be that for a static string in
Minkowski spacetime with a Gaussian perturbation to either the $X$ or
$P$ values. The evolution is then computed using the equations for a
dynamical string in a Minkowskian background [equations (85) and (86)
of paper I]. In both cases we find the same frequencies, to within an
amount $\Delta f = 0.01$, that we found in the original case of the
fully coupled system excited by a Weber--Wheeler pulse. Furthermore the 
frequencies did not depend on the location or shape of the field perturbation
nor upon the choice of $X$ or $P$ as the perturbed field.

\subsection{The long term behavior of the dynamic string}

\begin{figure}[t]
  \centering
  \epsfig{file=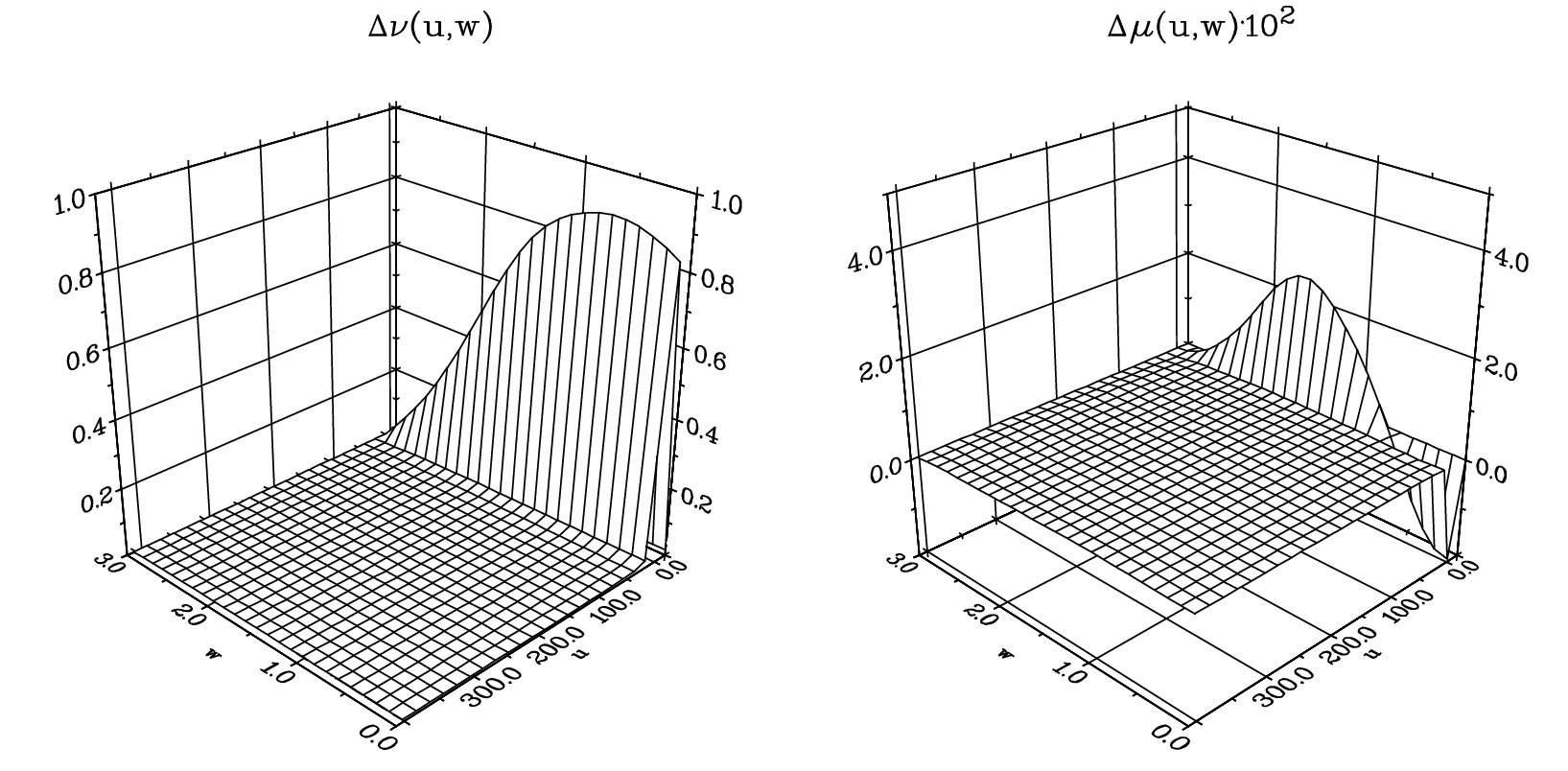, height=400pt, width=150pt, angle=-90}
  \epsfig{file=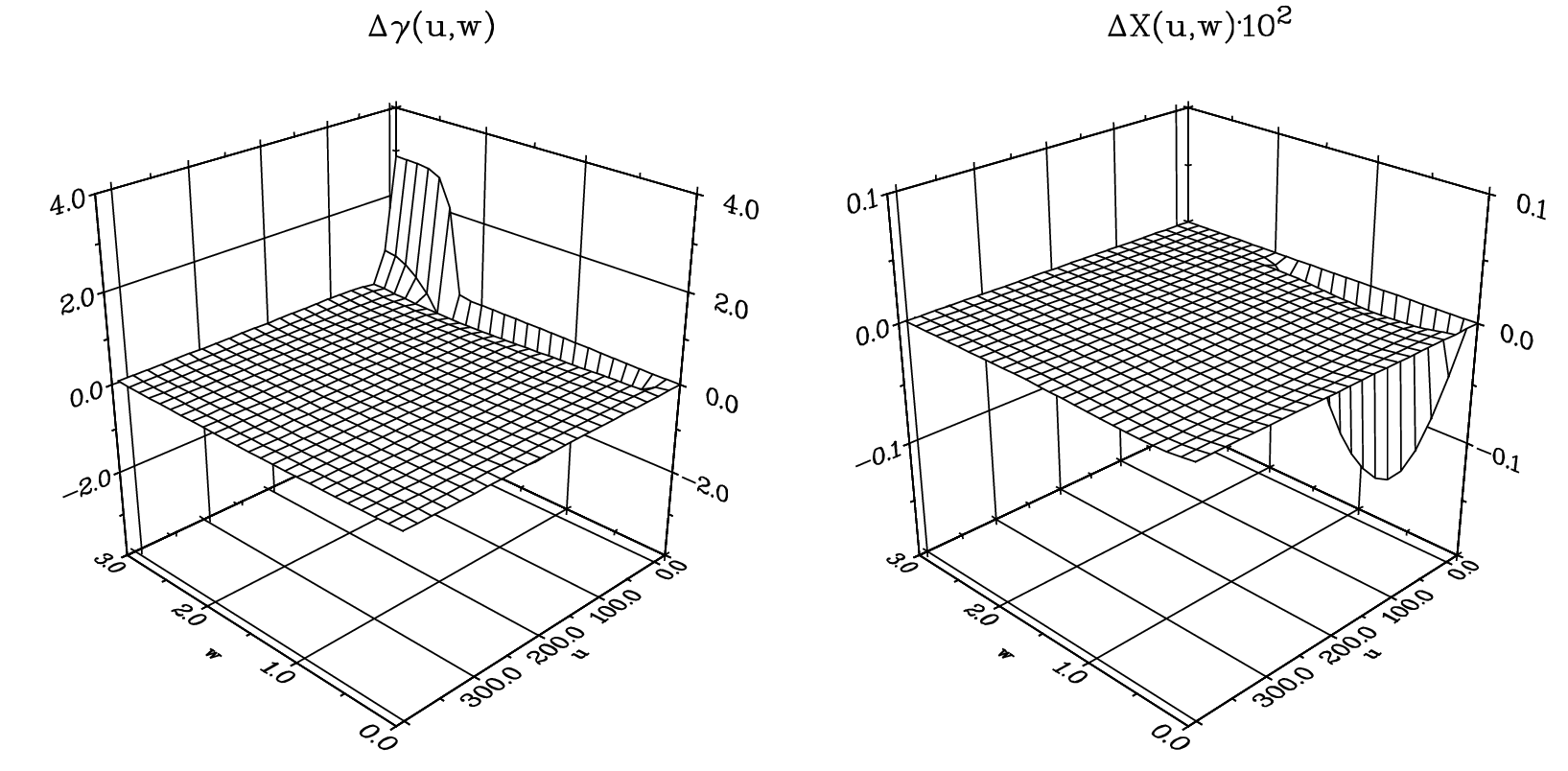, height=400pt, width=150pt, angle=-90}
  \epsfig{file=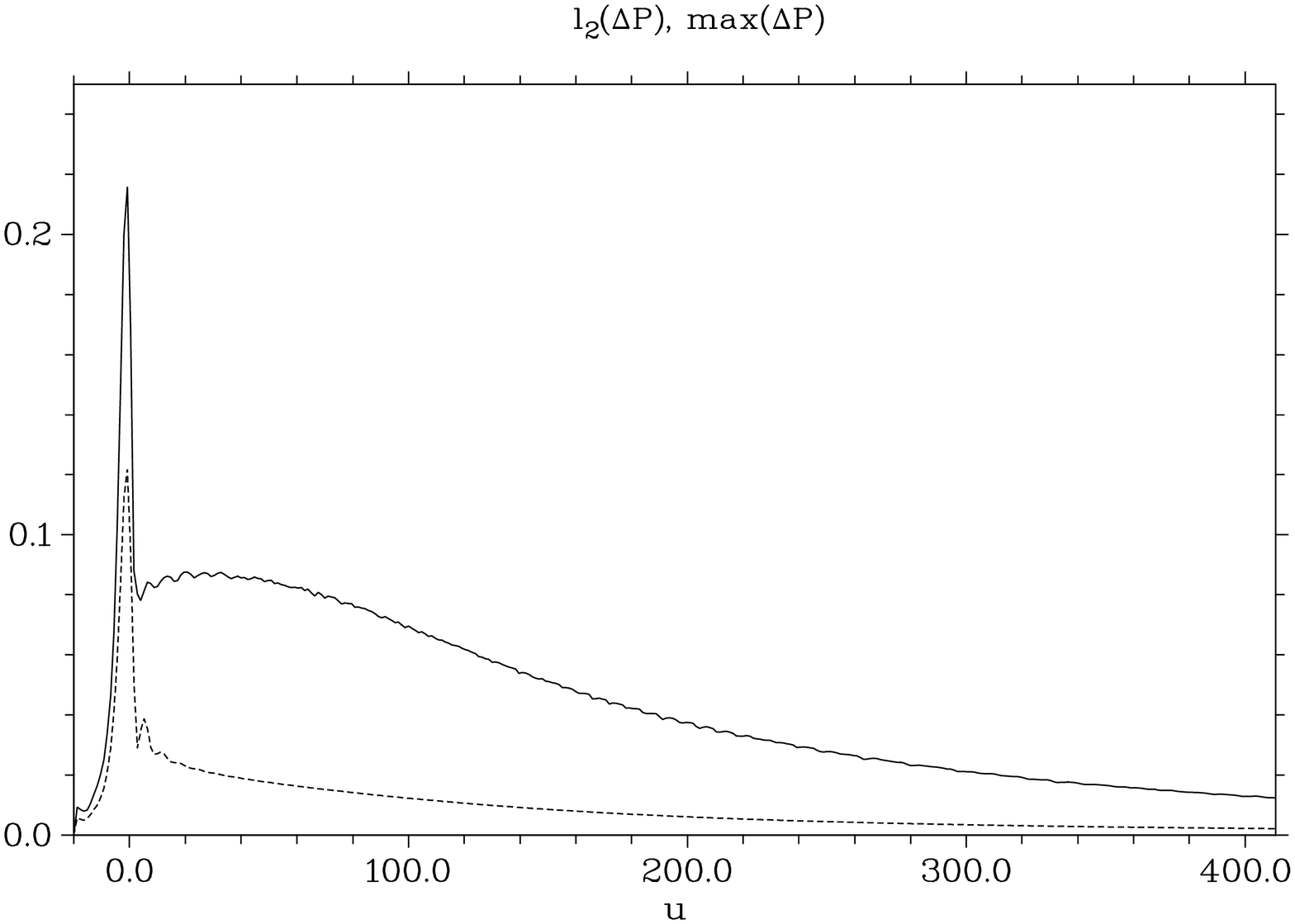, height=250pt, width=150pt, angle=-90}
  \caption{The upper four plots show the difference between the evolved
           functions $\nu$, $\mu$, $\gamma$ and $X$ and their corresponding
           static results. For $P$ a similar 3-dimensional plot is not suitable 
           since it fails to resolve the oscillations of the vector field. 
           Therefore we plot the $\ell_2$-norm (dashed line)
           and the maximum (solid line) of $\Delta P$ as 
           a function of time. $\nu$, $\mu$, $\gamma$ and $X$ quickly settle 
           down in their equilibrium configuration to numerical accuracy.
           The decay of the oscillations of $P$ takes significantly more
           time but eventually $P$ also approaches its equilibrium state.}
\label{settle}
\end{figure}

The time evolution shown in Figures \ref{mplot} and \ref{ringing} indicate that
the oscillations of the cosmic string excited by gravitational waves
gradually decay and metric and string settle down into an equilibrium state.
We have calculated a very long run
($-20\le u \le 410$) to investigate the long term behavior in detail. 
The unphysically large value of $\eta = 0.1$ is chosen for this
calculation in order to guarantee a
strong interaction between spacetime geometry and the cosmic string.
In Figure \ref{settle} we show the difference $\Delta f :=
f_{\rm evol} - f_{\rm stat}$
between the time-dependent $\nu$, $\mu$, $\gamma$ and $X$ and their 
corresponding static results obtained for the same parameters.
For the vector field $P$ a similar 3-dimensional plot would require
an extreme resolution to properly display the oscillations of the vector
field (cf.\ Figure \ref{mplot}). For this reason we proceed differently 
and calculate the $\ell_2$-norm and the maximum of $\Delta P$ for each
slice $u={\rm const}$. Both functions
are plotted versus time in Figure \ref{settle}. 
The incoming wave pulse can clearly be seen as a strong deviation of $\nu$
from the static function. The pulse excites the cosmic
string and is reflected at the origin at $u=0$. The metric variables and
the scalar field $X$ then quickly reach their equilibrium values. The
oscillations in $P$ decay on a significantly longer time scale 
which is also evident in Figures \ref{mplot} and \ref{ringing}
and the $\ell_2$-norm of $\Delta P$ slowly approaches zero.
Significantly longer runs than shown here are prohibited by the required
computational time, but the results indicate that $P$ will also eventually
reach its equilibrium configuration.

\section{Conclusion}
In this paper we have described the details of the implicit,
fully characteristic, numerical scheme which is used to solve the
field equations for a cosmic string coupled to gravity. A feature of
the cosmic string equations is that they admit exponentially diverging
unphysical solutions. By using a Geroch decomposition it is possible
to reformulate the problem in terms of fields which describe the
string on an asymptotically flat $2+1$-dimensional metric as well as
two auxiliary fields $\nu$ and $\tau$ which describe the
gravitational degrees of freedom. We can then introduce a conformally
compactified radial coordinate $y$ which allows us to include null
infinity as part of the numerical grid. As well as avoiding the need
to introduce artificial outgoing boundary conditions at the edge of
the grid this approach has the advantage that we can also enforce boundary
conditions for the string variables at null infinity which rule out
the unphysical solutions. The use of the geometrically defined Geroch
variables also improves the long term stability of the code compared
to the use of metric variables.

The code has been shown to reproduce the results of two exact vacuum solutions,
the Weber--Wheeler solution which describes a pulse of gravitational
radiation with just the $+$ polarisation state, and a solution due to
Xanthopoulos which describes a gravitational wave with both the $+$ and
$\times$ polarisation states. The code has also been shown to
reproduce the results of the static cosmic string code in that initial
data corresponding to a static solution do not change when
evolved in time using the dynamical code. For both the exact vacuum
solutions and the static initial data the code shows excellent long
term stability. Finally a time dependent convergence analysis
demonstrates clear second order convergence of the code.

After demonstrating the reliability of the code we use it to analyse
the interaction between an initially static cosmic string and a
Weber--Wheeler type pulse of gravitational radiation. We find that the
gravitational wave excites the string and causes the string variables
$X$ and $P$ to oscillate before the configuration slowly settles back into
its equilibrium state. In terms of the unphysical rescaled
variables we find that the frequencies of the oscillations are essentially
independent of the value of the coupling constant $\eta$ and of the
width and amplitude of the Weber--Wheeler pulse. We also find that the
frequency of $X$ is independent of $\alpha$ while that of $P$ is
proportional to $\sqrt \alpha$. When this result is translated back
into the physical units we find that the frequency of the scalar field
is proportional to the mass of the scalar field and the
frequency of the vector field is proportional to the mass of
the vector field. This result is confirmed by investigating two
further scenarios. Firstly we consider the evolution of static initial
data for the string coupled to the gravitational field, but with a
Gaussian perturbation to one of the string variables, and secondly we
consider the same thing but in a Minkowskian background with the
gravitational field decoupled. In both cases we obtain the same
relationship between the frequencies and the mass.

Having investigated the interaction between a Weber--Wheeler type pulse
of gravitational radiation and the cosmic string, the next obvious
step is to consider the interaction between the string and a pulse of
gravitational radiation with both polarisation states present. The
code will also be a valuable tool in comparing the numerical results with those
that one gets from using perturbation theory or the thin string limit
to approximate the behavior  of a cosmic string coupled to gravity.

\acknowledgements
We would like to thank Ray d'Inverno for helpful discussions and
Denis Pollney for help with GRTensor II.


\end{document}